\documentclass[12pt]{article}
\usepackage[utf8]{inputenc}
\usepackage{amsthm,amsmath,amssymb,graphicx,textcomp,gensymb} 
\usepackage{natbib}
\RequirePackage[colorlinks,citecolor=blue,urlcolor=blue]{hyperref}
\usepackage[dvipsnames]{xcolor}
\usepackage{color}
\usepackage{bm,bbm}
\usepackage[margin=2.4cm]{geometry}
\usepackage[skip=10pt plus1pt, indent=0pt]{parskip}
\usepackage{caption}
\usepackage{subcaption}
\usepackage{multirow}
\usepackage{dcolumn}
\allowdisplaybreaks

\title{\Large{Accounting for missing data when modelling block maxima}}
\author{Emma S.\ Simpson and Paul J.\ Northrop\\ 
\normalsize{Department of Statistical Science, University College London,}\\ \normalsize{Gower Street, London, WC1E 6BT, U.K.}\\
\normalsize{Email: \url{emma.simpson@ucl.ac.uk}, \url{p.northrop@ucl.ac.uk}}}
\date{}

\def\edit#1{{\textcolor{black}{#1}}}

\begin{document}
\maketitle

\begin{abstract}
Modelling block maxima using the generalised extreme value (GEV) distribution is a classical and widely used method for studying univariate extremes. It allows for theoretically motivated estimation of return levels, including extrapolation beyond the range of observed data. A frequently overlooked challenge in applying this methodology comes from handling datasets containing missing values. In this case, one cannot be sure whether the true maximum has been recorded in each block, and simply ignoring the issue can lead to biased parameter estimators and, crucially, underestimated return levels. We propose an extension of the standard block maxima approach to overcome such missing data issues. This is achieved by explicitly accounting for the proportion of missing values in each block within the GEV model. Inference is carried out using likelihood-based techniques, and we propose an update to commonly used diagnostic plots to assess model fit. We assess the performance of our method via a simulation study, with results that are competitive with the ``ideal’’ case of having no missing values. The practical use of our methodology is demonstrated on sea surge data from Brest, France, and air pollution data from Plymouth, U.K.
\end{abstract}


\noindent{\bf Code availability:} Code to carry out all methodology presented in this paper, including the simulation study and data examples, is available through the \texttt{R} package \texttt{evmiss} \citep{evmiss}. We provide a demonstration of the package through a supplemental \texttt{R} code file, which reproduces the plots presented here.

\noindent{\bf Data availability:} The Brest sea surge data studied in Section~\ref{subsec:seasurge} are available in the \texttt{R} package \texttt{Renext} \citep{Renext}, with the original source being \url{https://data.shom.fr}. The air pollution data considered in Section~\ref{subsec:ozone} were provided by the Department for Environment Food and Rural Affairs (Defra)\footnote{\copyright~Crown 2025 copyright Defra via \href{https://uk-air.defra.gov.uk}{uk-air.defra.gov.uk}, licensed under the \href{https://www.nationalarchives.gov.uk/doc/open-government-licence/version/2/}{Open Government Licence} (OGL).}, and downloaded via their Data Selector tool (\url{https://uk-air.defra.gov.uk/data/data_selector}). Both datasets are available as supplementary material in the \texttt{R} package \texttt{evmiss}.

\noindent{\bf Conflicts of interest:} The authors confirm that they have no conflicts of interest.

\newpage
\section{Introduction}\label{sec:intro}
Accurate statistical modelling plays a crucial role in understanding extreme events, allowing us to better prepare for and mitigate their impact. A wide range of examples arise in environmental applications, from studying weather-related events like heatwaves and droughts, to considering the health impacts of air pollution. The field of extreme value analysis provides a wealth of theoretically justified techniques that can be used in such settings, and importantly, allows for prediction at levels not previously observed in data.

In the univariate setting, a classical and widely used approach is to model the observed maximum value in each of a series of blocks. These blocks are often taken to be consecutive years, resulting in a sequence of annual maxima to be studied. Extreme value theory dictates that an appropriate model for these block maxima is the generalised extreme value (GEV) distribution. As well as providing insight into the behaviour of extreme events, this method can be used in the estimation of return levels, i.e., a value that is expected to be exceeded once in a specified time period, which corresponds to a particular quantile of an appropriate GEV distribution. We provide further statistical details on this approach in Section~\ref{sec:GEV}. 

While the technique of modelling block maxima using the GEV distribution is often criticised as being wasteful of data, since only one observation in each block contributes to the statistical analysis, \edit{there are various application areas where it is routinely implemented, such as in climate science \citep{Philip2020, Otto2024} and hydrology \citep{Yan2021,Sampaio2021}}. Moreover, in a recent comparison with the main alternative peaks-over-threshold approach, \cite{Bucher2021} show that there are cases where the block maxima method is actually preferred. However, one practical concern is how to proceed when faced with a dataset affected by missingness. This problem is often overlooked in practice, despite being commonly encountered in environmental settings, but can have serious consequences on the reliability of results if it is ignored. In particular, if some of the data in a given block are missing, it is not possible to know whether its true maximum value has actually been recorded. That is, a block maximum value extracted under missingness will be less than or equal to the true block maximum; this has the potential to bias our statistical analysis, including the underestimation of return levels. We provide an example of this in Section~\ref{subsec:missingnessExample}, as further motivation.

While the issues surrounding missing data when modelling block maxima have been acknowledged in previous literature, attempts to deal with these often appear to be somewhat arbitrary. \edit{Until recently,} approaches have been mostly limited to discarding blocks where the proportion of missingness is deemed ``too high''; see, for example, \cite{Vandeskog2022} who use a \emph{blended} GEV model for precipitation maxima. While this is a reasonable approach that may reduce bias compared to ignoring the issue completely, results may be sensitive to the level of missingness that is deemed acceptable, and some negatively biased block maxima will still remain. In addition, removing some of the available blocks due to missingness will only exacerbate the issue of data scarcity that is commonly associated with block maxima modelling, thus potentially leading to increased estimation uncertainty. \cite{Hossain2022} also mention interpolation and spatial pooling as possible solutions to the problem of missing data in block maxima modelling, but do not provide much specific detail on their procedure. Time series interpolation is common in the general missing data literature, but this approach is potentially problematic in the context of extreme value modelling since interpolated values are restricted to the range of the available (non-missing) data. It is therefore unlikely to correct the block maxima values, and the same return level underestimation issues would persist. On the other hand, spatial pooling is a potentially reasonable approach, but only in situations where appropriate data are available, which is by no means a given. \edit{Independently and contemporaneously to this work, \cite{McVittie2025a,McVittie2025b} have also recently considered the challenge of handling missing data when modelling block maxima. They are motivated by modelling extreme wave surges, and propose estimation of the usual GEV parameters through censored and weighted likelihood techniques; we provide further detail on the latter approach in Section~\ref{sec:simulations}.}

The issues discussed above provide the motivation for this paper. Our aim is to develop a new approach to GEV model fitting in the presence of missing data, that avoids relying on the availability of supplemental datasets. Our method makes a simple adjustment to the usual GEV distribution by taking into account the proportion of missing values per block. This avoids the need to discard any information or make subjective decisions about how much missingness is acceptable, while still providing a robust approach to parameter and return level estimation. We do emphasise that we work in the rather idealised setting of independent and identically distributed (i.i.d.)\ data with non-informative missingness, but believe our approach to be a reasonable first step in tackling this important problem. We further discuss the limitations and potential extensions of our proposed methodology in Section~\ref{sec:discussion}.

To summarise the contents of the remainder of the paper, we begin in Section~\ref{sec:GEV} by providing an overview of the standard approach to using the GEV distribution to model block maxima. Section~\ref{sec:GEVmiss} details our proposed extension of the GEV model for block maxima that are affected by missing data, and provides a strategy for assessing model fit. In Section~\ref{sec:simulations}, we demonstrate the performance of our approach through a simulation study, \edit{with comparison to some competing estimators,} while in Section~\ref{sec:application} the method is applied to two environmental datasets related to sea surges and air pollution. Section~\ref{sec:discussion} concludes with a discussion of limitations and possible future work.

\section{The GEV distribution for block maxima modelling}\label{sec:GEV}

In this section, we provide an overview of the main ideas around using the GEV distribution for modelling block maxima. These ideas date back to the work of \cite{FisherTippett1928}, \cite{vonMises1936} and \cite{Gnedenko1943}, and have been widely adopted for modelling univariate extremes. For more detail, and a popular textbook treatment of these ideas, we refer the reader to \cite{Coles2001}. 

\subsection{The extremal types theorem}\label{sec:ETT}
Suppose we have $n$ i.i.d.\ random variables $X_1,\dots,X_n$ \edit{with common distribution function $F_X(\cdot)$}, and consider their maximum, denoted by
$M_n = \max\{X_1,\dots,X_n\}$. The extremal types theorem \citep[see][]{Leadbetter1983} states that if the limiting distribution of a suitably standardised version of $M_n$ is non-degenerate, this must belong to a particular class of models, known as the GEV distribution \citep{Jenkinson1955}. That is, we assume that there exist standardising functions, $a(n)$ and $b(n)>0$, such that
\begin{align}
    \Pr\left\{\frac{M_n - a(n)}{b(n)}\leq z\right\}\rightarrow  
G(z;\mu,\sigma,\xi), \qquad \text{as $n\rightarrow\infty$,}
    \label{eqn:GEVdefinition}
\end{align}
where $G(z;\mu,\sigma,\xi)$ is the distribution function of a GEV$(\mu,\sigma,\xi)$ distribution. This has the form
\begin{align}
    G(z;\mu,\sigma,\xi) = \begin{cases}
        \exp\left[-\left\{1+\xi\left(\frac{z-\mu}{\sigma}\right)\right\}_+^{-1/\xi}\right], &\xi\neq 0,\\
        \exp\left[-\exp\left\{-\left(\frac{z-\mu}{\sigma}\right)\right\}\right], &\xi=0,
    \end{cases}
    \label{eqn:GEVdistn}
\end{align}
for $t_+=\max(0,t)$ and with $\mu\in\mathbb{R}$, $\sigma>0$ and $\xi\in\mathbb{R}$ termed the location, scale and shape parameters, respectively.

In general, considering $n$ tending towards infinity, as in~\eqref{eqn:GEVdefinition}, is not of practical use. Instead, this limiting result is commonly taken to be an approximation for \textit{large enough} values of $n$. For a fixed value of such $n$, the standardisation of $M_n$ in~\eqref{eqn:GEVdefinition} can be ``undone'', leading instead to the assumption that
\begin{align}
 M_n\sim \text{GEV}\left(a(n) + \mu b(n), \sigma b(n), \xi\right).
 \label{eqn:n-dependentGEV}
\end{align}
Since $n$ itself is fixed here, the standardising functions $a(n)$ and $b(n)$ appearing in the updated location and scale parameters can be ignored, and we can simply assume that 
\[
M_n\sim \text{GEV}\left(\mu_n , \sigma_n, \xi\right), \qquad \text{for some $\mu_n\in\mathbb{R}$, $\sigma_n>0$ and $\xi\in\mathbb{R}$}.
\]
For simplicity, we remove the $n$ subscript on the GEV parameters in the following, letting $M_n\sim \text{GEV}\left(\mu, \sigma, \xi\right)$ with $\mu\in\mathbb{R}$, $\sigma>0$ and $\xi\in\mathbb{R}$ for a specified (large) value of $n$.

\subsection{Block maxima modelling}
Statistical inference for maxima using the GEV distribution requires multiple observations of the random variable $M_n$. Suppose that we start with $n\times b$ observations corresponding to the underlying $X$ variables, which we denote by $x_1,\dots,x_{nb}$, often referred to as the \textit{raw data}. The standard approach is to separate these values into $b$ consecutive and non-overlapping blocks, each of length $n$, and to consider the maximum in each one. That is, to define
\[
m_i = \max\left\{x_{(i-1)n+1},\dots,x_{in}\right\},\qquad \text{for $i=1,\dots,b$},
\]
where the values $m_1,\dots,m_b$ are collectively referred to as the \textit{block maxima}. 

For environmental applications, it is common to take block lengths of one year. For example, where the original data are measured on a daily scale, this corresponds to having $n=365$ (or $n=366$ for leap years). The reason for this is that such data often exhibit seasonality, but this can be removed in the process of taking annual block maxima, with the GEV assumption for these values often still being reasonable. If annual blocks are not long enough for the asymptotic results in~\eqref{eqn:GEVdefinition} to hold, one may consider using multiple years in each block, with the trade-off that this reduces the value of $b$, thereby increasing estimation uncertainty. In this case, the break points between blocks should also be carefully chosen so that the maxima generally occur towards the centre of each block, reducing the chance of some block maxima being dependent.

Given observations of the block maxima, the parameters of the GEV distribution can be estimated using a wide range of techniques, such as via maximum likelihood estimation or Bayesian approaches.

\subsection{Return levels}
The purpose of carrying out block maxima modelling is usually to assess the behaviour or occurrence of extreme values, often at levels beyond those previously seen. A useful quantity here is the \textit{return level}, which can be thought of as the value that is expected to be exceeded once in a specified number of blocks (referred to as the \textit{return period}). Suppose we are interested in a return period corresponding to $r$ blocks, denoting the corresponding return level by $z_r$. The $r$-block return level $z_r$ has probability $1/r$ of being exceeded in any single block and is therefore the $(1-1/r)$ quantile of the GEV distribution of interest, i.e.,
\begin{align}
    z_r = 
    \begin{cases}
        \mu - \frac{\sigma}{\xi}\left[1-\left\{-\log\left(1-\frac{1}{r}\right)\right\}^{-\xi}\right], &\xi\neq 0,\\
        \mu - \sigma\log\left\{-\log\left(1-\frac{1}{r}\right)\right\}, &\xi=0.
    \end{cases}
    \label{eqn:RLs}
\end{align}

Return level estimates can be obtained by replacing the GEV model parameters $(\mu,\sigma,\xi)$ in~\eqref{eqn:RLs} by their estimated values $(\hat\mu,\hat\sigma,\hat\xi)$. In a frequentist setting, an equivalent approach is to profile the log-likelihood with respect to $z_r$, and maximise the resulting function directly to estimate the required return level \citep[see Section 3.3.4 of][]{Coles2001}. Profiling is often preferred over standard likelihood inference due to its more reliable estimation of confidence intervals.

\section{Extension of the GEV to handle missing data}\label{sec:GEVmiss}
\subsection{\edit{An illustration} of the missing data issue}\label{subsec:missingnessExample}
\begin{figure}[!b]
    \centering
    \includegraphics[width=0.9\textwidth]{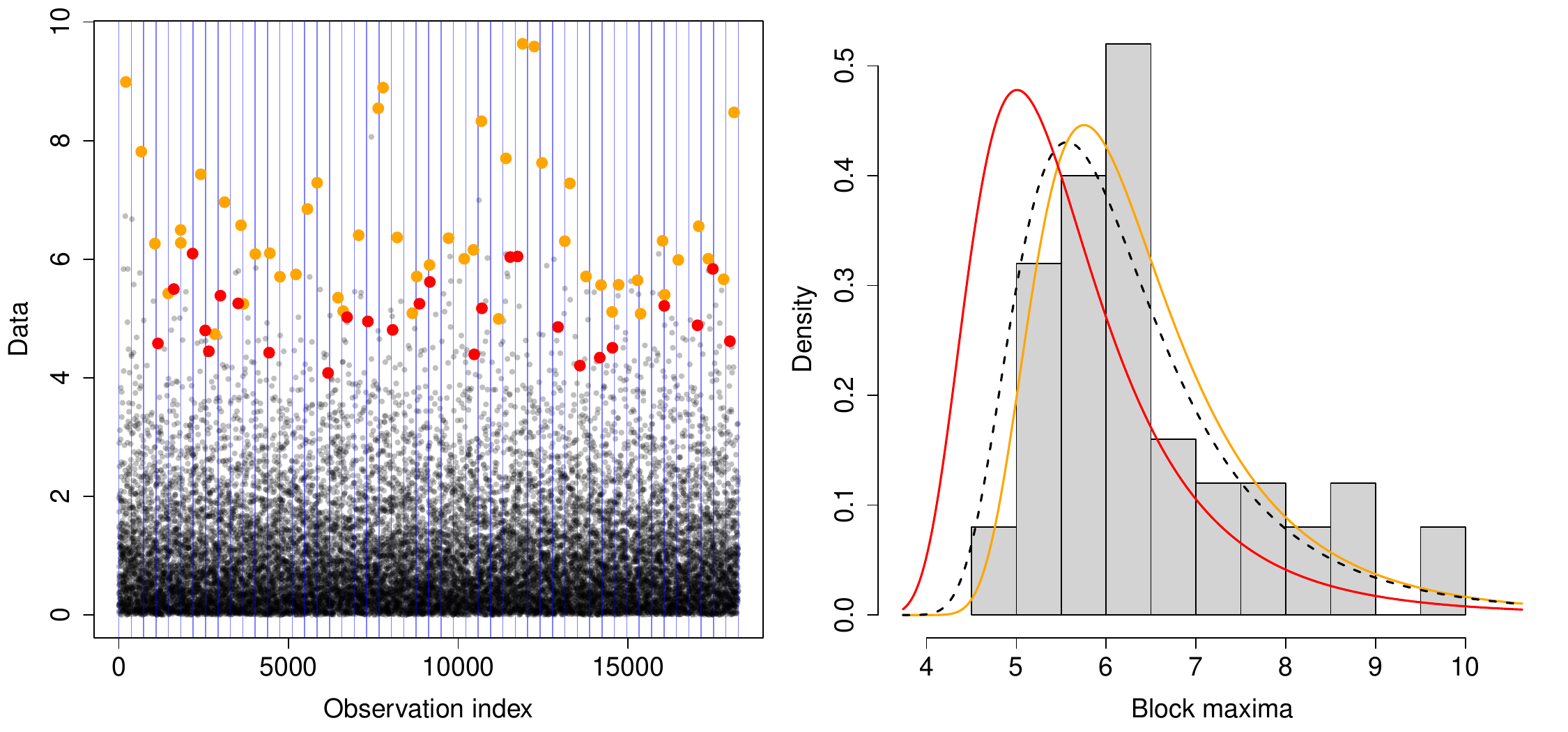}
    \caption{Left: Simulated data with no missingness (grey points) separated into blocks of length $n=365$ with the corresponding block maxima highlighted in orange. The red points show the block maxima that decrease once missingness is introduced. Right: Histogram of the original block maxima, with the fitted GEV densities for the full data (orange), and block maxima under missingness (red). The black dashed line shows the estimated density under missingness using our new approach; this result will be discussed further in Section~\ref{subsec:diagnostics}.}
    \label{fig:blockmax}
\end{figure}
To emphasise the need to carefully consider the handling of missing data in the context of block maxima modelling, we begin with an example using the standard GEV approach described in Section~\ref{sec:GEV}. 

We simulate raw data from a standard exponential distribution, taking $b=50$ blocks, each of length $n=365$; these data are shown in the left panel of Figure~\ref{fig:blockmax}. The block maxima are extracted (shown in orange), and a GEV distribution is fitted, resulting in the estimated GEV density shown in orange in the right panel of Figure~\ref{fig:blockmax}. The estimated GEV parameters in this case are $(\hat\mu, \hat\sigma, \hat\xi)=(5.87,0.83,0.15)$. We then randomly remove 50\% of the raw data and recalculate the 50 block maxima under missingness. The resulting GEV fit for these missingness-affected block maxima is shown in red in Figure~\ref{fig:blockmax}, with the GEV parameter estimates now being $(\hat\mu, \hat\sigma, \hat\xi)=(5.12,0.78,0.15)$.

Under missingness, it is clear that the negative bias in the block maxima values has resulted in a biased GEV fit, which in this case particularly transpires through the location parameter. This would lead to underestimation of the corresponding return levels, and demonstrates the risk of ignoring missing data when modelling block maxima. In the remainder of this section, we propose an approach to account for this missingness. 

\subsection{General strategy}\label{subsec:generalApproach}
Let $n$ denote the maximum block size, i.e., the block size under no missingness, and recall that $b$ denotes the number of available blocks. We consider the underlying random variables $X_1,\dots,X_{nb}$ to be i.i.d., so that with large enough $n$ and no missingness, we make the standard assumption that $M_n\sim\text{GEV}\left(\mu, \sigma, \xi\right)$ with distribution function $G(z;\mu,\sigma,\xi)$ as in~\eqref{eqn:GEVdistn}. 

Now let $n_i\leq n$ denote the number of non-missing observations in the $i$th block, for $i=1,\dots,b$. We assume here that observations are missing completely at random in each block. The missingness mechanism can vary between blocks, but must be non-informative, meaning that whether or not an observation is missing does not depend on its value. Motivated by assumption~\eqref{eqn:n-dependentGEV}, our general approach for handling missingness in modelling block maxima is to allow the GEV location and scale parameters to depend on the $n_i$ values, i.e., to let
\begin{align}
M_{n_i}\sim\text{GEV}\left(\mu(n_i) , \sigma(n_i), \xi\right), \qquad i=1,\dots,b,
\label{eqn:GEVmissassumption}
\end{align}
for some functions $\mu:\mathbb{Z}_+\rightarrow\mathbb{R}$ and $\sigma:\mathbb{Z}_+\rightarrow\mathbb{R}_{>0}$, and $\xi\in\mathbb{R}$. \edit{In practice, this approach requires that the number of non-missing observations per block is known, e.g., through access to the raw data with missing data flags.}

Considering the range of possible standardising functions that can arise in assumption~\eqref{eqn:GEVdefinition} \citep[for examples, see Section~3.1.5 of][]{Coles2001}, one possibility is to impose flexible, non-linear, parametric forms on $\mu(\cdot)$ and $\sigma(\cdot)$, e.g., by exploiting Box-Cox functions. This allows us to treat the missing data problem as a regression task, with the number of non-missing values as a covariate. In our investigations, we found this to be a promising approach, but one that had some drawbacks. First, for identifiability of the parameters, we need to observe a range of missingness proportions across blocks, which is not always guaranteed. Additionally, the act of taking block maxima intrinsically leads to a limited number of observations; it is not uncommon in environmental applications to have time series of around 30-50 years, and increasing the number of model parameters makes estimation a more difficult task. Finally, we found estimation to be much slower computationally for these regression-type models, compared to the standard GEV approach, which is a downside if they are to be adopted in practice.

It may have been possible to refine the above method to address the issues highlighted, but this comes at the risk of overcomplicating the approach. Instead, we propose an alternative method that avoids the introduction of additional model parameters while still accounting for the amount of missingness in each block. 

\subsection{A more parsimonious approach}\label{subsec:parsimony}
As a more parsimonious solution to the missing data problem, we now propose to infer an approximate distribution for each $M_{n_i}$ directly from $\{G(z;\mu,\sigma,\xi)\}^{n_i/n}$. \edit{This approximation allows us to exploit the max-stability property of the GEV distribution. Further intuition is provided by considering that for large $n$, we have
\[
\Pr\left(M_n\leq z\right) = \Pr\left(X_1 \leq z,\dots,X_n \leq z\right) = F_X(z)^n \approx G(z;\mu,\sigma,\xi),
\]
so that when only $n_i$ observations are available in the $i$th block, we should instead consider}
\[
\edit{\Pr(M_{n_i}\leq z) = F_X(z)^{n_i} = \left\{F_X(z)^{n}\right\}^{n_i/n}\approx \{G(z;\mu,\sigma,\xi)\}^{n_i/n}.}
\]
\edit{We demonstrate in Section~\ref{sm:GEVmissingParameters} of the Supplementary Material} that this assumption is equivalent to having $M_{n_i}$ follow a GEV distribution as in~\eqref{eqn:GEVmissassumption}, but with missingness-dependent location and scale parameters taking the specific forms
\begin{align}
\mu(n_i) &= \begin{cases}
        \mu+\frac{\sigma}{\xi}\left\{\left(\frac{n_i}{n}\right)^\xi-1\right\}, &\xi\neq 0,\\
        \mu + \sigma\log(n_i/n), &\xi=0,
    \end{cases}\nonumber\\
    \sigma(n_i) &= \sigma \left(\frac{n_i}{n}\right)^\xi.
\label{eqn:locscale_miss}
\end{align}
To reiterate, this model has a benefit over the general approach outlined in Section~\ref{subsec:generalApproach}, in that it involves only three parameters. It is therefore no more complicated than the standard GEV model, but allows the level of missingness to be appropriately accounted for in block maxima modelling. We note that the location parameter in~\eqref{eqn:locscale_miss} takes a Box-Cox-type form, with $\mu(n_i) = \mu + \sigma BC(n_i / n, \xi)$, where $BC(\cdot,\xi)$ is the one-parameter Box-Cox transformation function with parameter $\xi$, highlighting a further link with the more general approach presented above.

We propose to carry out estimation using standard maximum likelihood techniques, with profiling used where appropriate. Return level estimates can be obtained by considering relevant quantiles of the corresponding $\text{GEV}(\mu,\sigma,\xi)$ distribution (i.e., setting $n_i=n$) in~\eqref{eqn:locscale_miss}. Despite the simplicity of this approach, in the simulation study of Section~\ref{sec:simulations}, we show it to be competitive with results in the ideal case where all data are available, and to provide much improvement over the na\"ive approach of ignoring missingness completely.

\subsection{Diagnostic plots}\label{subsec:diagnostics}
For model checking, we propose to adapt the visual diagnostics provided for the usual GEV distribution in the \texttt{R} package \texttt{ismev} \citep{ismev}, as described by \cite{Coles2001}. These include four plots, namely a PP-plot, a QQ-plot, a return level plot and a density histogram. 

In our setting, it is straightforward to construct the PP-plot, since the model-based cumulative probability for the observed maximum in the $i$th block is simply
\[
    \hat p_i = G\left(m_i;\hat\mu(n_i),\hat\sigma(n_i),\hat\xi\right), \qquad i=1,\dots,b,
\]
where $m_i$ is the observed block maximum in the $i$th block and $(\hat\mu(n_i),\hat\sigma(n_i),\hat\xi)$ are the estimated GEV model parameters for a block with $n_i$ non-missing observations. Letting $\{\hat p_{(1)},\dots,\hat p_{(b)}\}$ denote an ordered version of the $\hat p_i$ values, i.e., where $\hat p_{(1)}\leq \hat p_{(2)}\leq\dots\leq\hat p_{(b)}$, the PP-plot consists of the points 
\[
    \left\{\left(\frac{i}{b+1}~,~\hat p_{(i)}\right): i=1,\dots,b\right\}.
\]

It is well documented that PP-plots can be unhelpful when studying extremes, since issues with the fit for the largest values are concealed. QQ-plots and return level plots overcome this issue. Under missingness, the observed block maxima are generally not identically distributed, so to construct the remaining plots we propose to first scale the observed block maxima to equivalent full-block maxima by matching quantiles of the relevant GEV distributions. Our adjusted block maximum for the $i$th block is
\begin{align*}
    \hat m^{\text{adj}}_i = G^{-1}\left(\hat p_i; \hat\mu(n),\hat\sigma(n),\hat\xi\right)= G^{-1}\left(\hat p_i; \hat\mu,\hat\sigma,\hat\xi\right),\qquad i=1,\dots,b.
\label{eqn:adjustedBM}
\end{align*}

Once the values of $\hat m_1^{\text{adj}},\dots,\hat m_b^{\text{adj}}$ are obtained, construction of the remaining diagnostic plots proceeds as usual. For the QQ-plot, we consider an ordered version of the adjusted block maxima, denoted by $\left\{\hat m^{\text{adj}}_{(1)},\dots,\hat m^{\text{adj}}_{(b)}\right\}$, with $\hat m^{\text{adj}}_{(1)}\leq \hat m^{\text{adj}}_{(2)}\dots\leq \hat m^{\text{adj}}_{(b)}$, and plot the points
\[
    \left\{\left(G^{-1}\left(\frac{i}{b+1}; \hat\mu,\hat\sigma,\hat\xi\right)~,~ \hat m^{\text{adj}}_{(i)}\right): i=1,\dots,b\right\}.
\]
To both the PP- and QQ-plots we add pointwise bands, given by the 2.5\% and 97.5\% quantiles of the relevant order statistic of the $\text{U}(0, 1)$ and fitted GEV distribution, respectively. \edit{For the QQ-plot, an alternative to standardising the observed block maxima to the full-block scale is to adjust the positions of the empirical points plotted on the $x$-axis, as in \cite{Belzile2022}. It is likely that similar conclusions would be drawn under the two methods, and we prefer to consider the standardised block maxima as these also facilitate production of the return level plot and density histogram below.}

\edit{For the return level plot, we follow a similar approach to \cite{Coles2001}, but with our horizontal axis representing $x=-\log(1-1/r)$ plotted on $-\log_{10}$-scale, and the vertical axis still showing the corresponding return levels in~\eqref{eqn:RLs}.} We make a small adaptation to the labelling used in \texttt{ismev} on the horizontal axis, showing selected values of the return period $r$ rather than $x$. This is approximately equivalent to the \texttt{ismev} approach for large $r$, but allows us to also accurately represent the return period for small $r$. In doing this, we preserve the feature that $\xi=0$ corresponds to the straight line, \edit{as in \cite{Coles2001}, }while providing a clearer link between the return periods and return levels across the full range of values. Profile-based $95\%$ \edit{pointwise asymptotic} confidence intervals are also added to the return level plot to aid comparison between the modelled and empirical results.

Finally, our density histogram is simply constructed from the points $\left\{\hat m^{\text{adj}}_i:i=1,\dots,b\right\}$, and a GEV density with parameters $(\hat\mu,\hat\sigma,\hat\xi)$ is superimposed, equivalent to the plot in the right panel of Figure~\ref{fig:blockmax}.

\begin{figure}[t]
    \centering
    \includegraphics[width=0.725\textwidth]{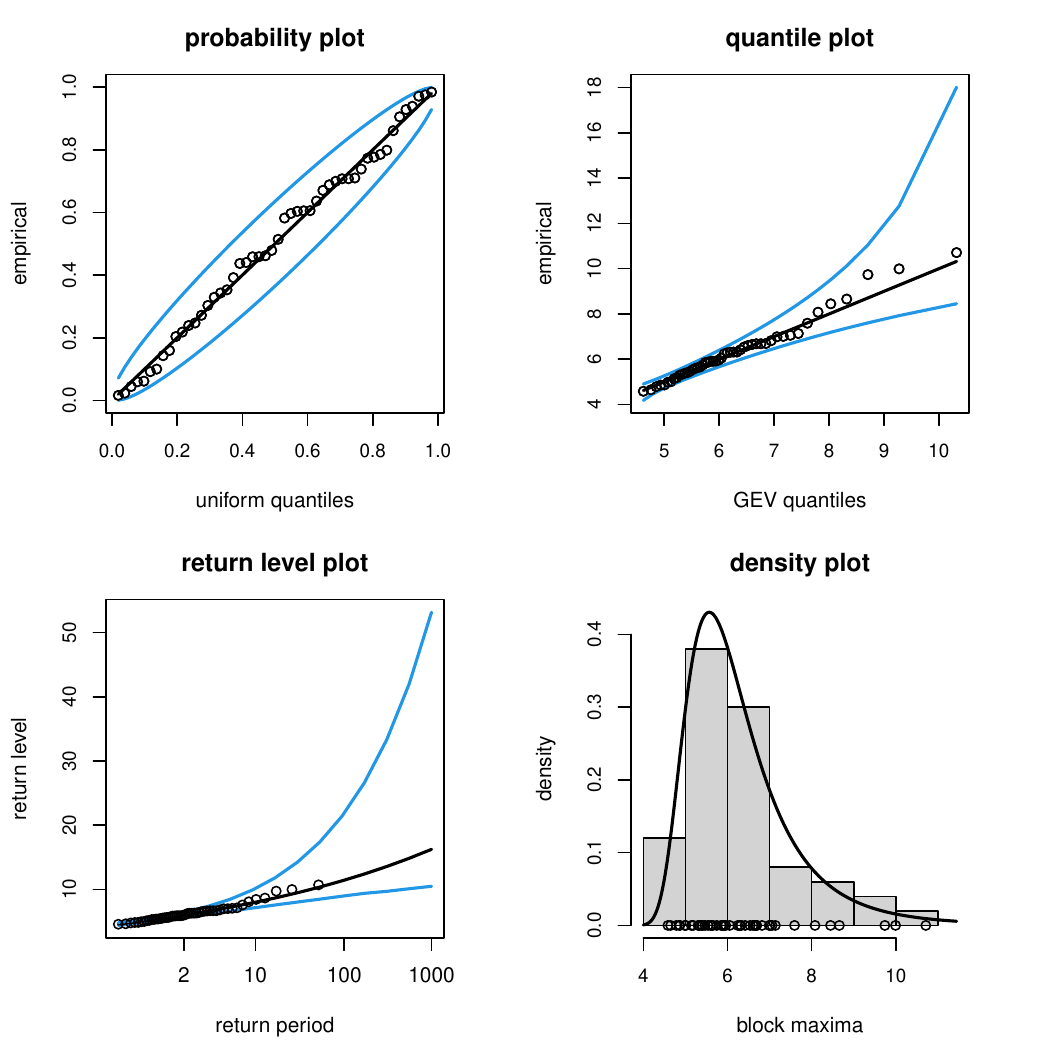}
    \caption{The four diagnostic plots, with adjustments for missingness, for the data in Figure~\ref{fig:blockmax}.}
    \label{fig:diagnosticExample}
\end{figure}

Interpretation of all four plots is done in the usual way. As an example, we fit our new model to the block maxima data with missingness from Figure~\ref{fig:blockmax}, obtaining parameter estimates $(\hat\mu,\hat\sigma,\hat\xi)=(5.69,0.86,0.15)$ and the diagnostic plots shown in Figure~\ref{fig:diagnosticExample}. The points on both the PP-plot and the QQ-plot lie close to the diagonal, with only slight deviations for the largest quantiles in the latter plot. These indications of a good model fit are supported by the return level plot, where all empirical points lie close to the modelled return level line and well within the associated confidence intervals. The estimated density also matches the shape of the histogram well. As a final check on the performance of our proposed method, we add the estimated density function to the right panel of Figure~\ref{fig:blockmax}. Clearly, we have been able to go a large way towards correcting the missingness-induced bias in this case. We provide a more thorough assessment of our approach in the simulation study of Section~\ref{sec:simulations}. Further diagnostic plot examples are provided for our data applications in Section~\ref{sec:application}.

\section{Simulation study}\label{sec:simulations}

\subsection{Simulation set-up}
We consider \edit{four} different distributions for the original variables $X_1,\dots,X_{nb}$. \edit{These are the standard exponential, the standard Gaussian, the Student's $t$ distribution with 2 degrees of freedom, and a Beta$(1,10)$ distribution.} For the first two choices, the convergence in~\eqref{eqn:GEVdefinition} leads to a GEV distribution with shape parameter $\xi=0$, while the \edit{third has $\xi=1/2$, and the final option has $\xi=-1/10$,} so that these distributions together allow us to study a range of different tail behaviours. In each iteration, we simulate $b=50$ blocks of length \edit{$n=90$, representing daily data from blocks corresponding to individual seasons, with a total length of time series that would reasonably be seen in practical applications.}

For our missingness mechanism, separately for each block, we generate a proportion of missingness from a \edit{$\text{U}(0,0.2)$ distribution. We then remove this proportion of observations from the block, completely at random. Overall, this results in around 10\% of the raw data being masked,} but with the proportion of missingness varying between blocks\edit{; this is again a realistic scenario, as reflected by the examples presented in Section~\ref{sec:application}}.

We apply our method proposed in Section~\ref{sec:GEVmiss} to estimate the GEV model parameters and 100-block return levels, along with their associated profile-based confidence intervals. These results are compared to \edit{four alternative approaches, details of which are provided in Section~\ref{subsec:SSmethods}. Each simulation setting is repeated 10,000 times, with two broad types of result presented in Section~\ref{subsec:SSresults}.} \edit{One type compares the GEV fit from each approach to a GEV distribution fitted to the full dataset before missingness was imposed (we refer to this method by the term ``full''). This comparison is, of course, impossible in practice, but, in the context of a simulation study, it allows us to make direct assessments against the ideal scenario of no missingness. For a finite block length, there are no exact true values of the GEV parameters. We use the penultimate approximation of \cite{Smith1987}, implemented in the \texttt{R} package \texttt{mev} \citep{mev_package}, to provide a guide to the parameter values, and hence return levels, that may be expected for a block length of 90. The other type of result compares inferences made using each approach to the known 100-block return level.}

\subsection{\edit{Methods for comparison}}\label{subsec:SSmethods}
\edit{The first two approaches to which we compare are ones that would currently often be seen in practice. The first is a na\"ive approach where the GEV distribution is fitted ignoring the missingness completely, and the second case sees blocks with more than 10\% of values missing discarded, before fitting the GEV distribution as usual with no other adjustment for missingness. We refer to these two approaches as ``na\"ive'' and ``discard'', respectively. We refer to our own approach by the name ``adjust''.\\
We also consider two estimators arising from recent research by \cite{McVittie2025b}, who use a weighted GEV log-likelihood, with the contribution from the maximum $m_i$ of the $i$th block multiplied by a weight $w_i$ that depends on $m_i$ and/or the number of non-missing values $n_i$. The first weighting scheme uses $w_i = n_i/n$ and the second $\hat{F}(m_i) ^ {n - n_i}$, where $\hat{F}$ is the empirical distribution function of the block maxima $m_1, \ldots, m_b$. We refer to these approaches as ``weight1'' and ``weight2'', respectively. For each weighting scheme, the larger the number $n - n_i$ of missing values, the smaller the weight.}

\subsection{Simulation results}\label{subsec:SSresults}

\begin{figure}[p]
    \centering
    \includegraphics[width=0.9\textwidth]{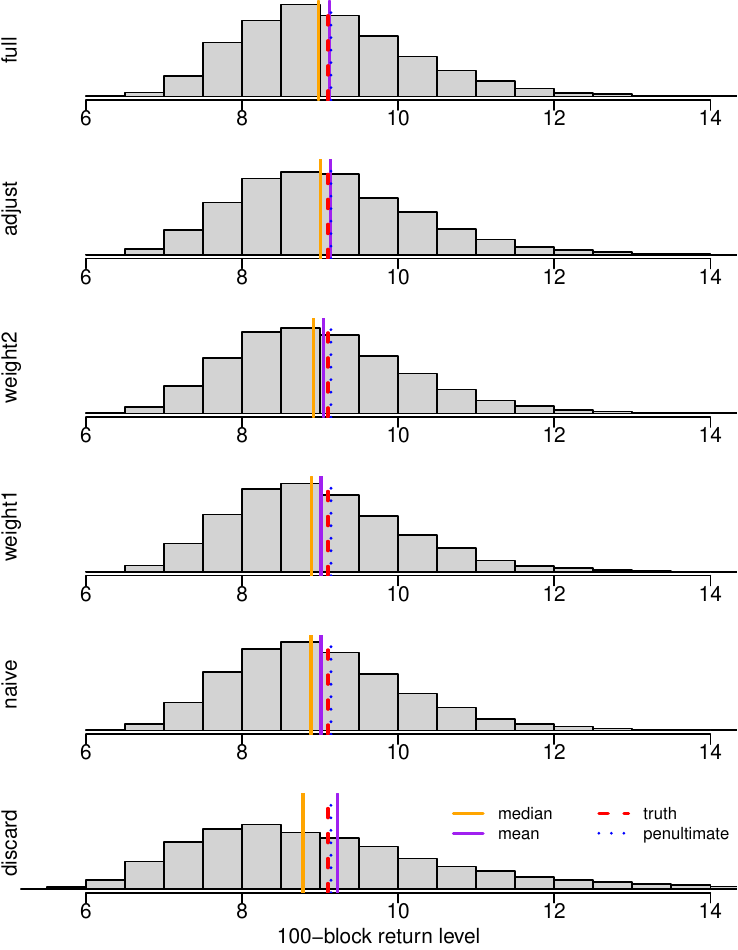}
    \caption{Histograms of estimated 100-block return levels across 10,000 simulation replicates for a standard exponential distribution. The vertical lines indicate the sample median and sample mean of the return level estimates, the true return level and the return level based on a penultimate approximation of the GEV parameters.}
    \label{fig:simulations_RLhistograms}
\end{figure}

Figure~\ref{fig:simulations_RLhistograms} depicts the sampling distributions of the estimators of the 100-block return level for each of the \edit{six approaches described above, and for underlying data simulated from a standard exponential distribution. Equivalent plots for the three other underlying distributions we consider are provided in Section~\ref{sm:figure3} of the Supplementary Material.} Superimposed on each plot are vertical lines indicating the true return level, the return level resulting from the penultimate approximation of the GEV parameters and the estimated mean and median of the given estimator. As is typical, the sampling distributions are positively skewed, strongly so for the Student's $t_2$ case, so it is instructive to consider both the mean and median as measures of average. These plots provide a summary of the main findings, supported by the numerical comparisons in the tables that follow. 

Our adjustment produces plots that are much like those based on the full dataset and the biases in estimation of the true return level are similar. Our adjusted estimator is less precise, with histograms showing a slightly greater spread of estimates with the median and mean lines being spread a little further apart. This is appropriate given the loss of information from removing observations. In the Student's $t_2$ case, both the ``full'' and ``adjust'' approaches produce an estimator that is approximately median unbiased, but the positive skewness of their sampling distributions result in means that lie above the true return level. 

As expected, the na\"ive estimator tends to underestimate relative to the ``full'' approach, but in the Student's $t_2$ case its mean is approximately equal to the truth. The estimates from the ``na\"ive'' approach vary less than those produced by our adjustment, but this apparent precision is misleading as it is based on the supposition that the maxima are from complete blocks of raw data. The main feature of the plot for the ``discard'' approach is that the estimates are more variable than the other approaches, owing to the reduced information in a sample, and exhibit stronger positive skewness, particularly in the Student's $t_2$ case, where the presence of some very large return level estimates causes the estimated mean of the estimator to be substantially larger than its estimated median. Moreover, for a small number of simulated datasets \edit{(13 for the exponential case, 22 for the Gaussian case and 46 for the beta case)} the search for the maximum likelihood estimator failed once block maxima had been discarded, so this is not always a sound approach.

\edit{The two weighted likelihood approaches of \cite{McVittie2025b} produce differing results. In general, their ``weight1'' approach produces results that are similar to the na\"ive approach of ignoring missingness completely. On the other hand, their ``weight2'' adjustment is reasonably successful, but in terms of median bias, is outperformed by our approach in all cases.}

The following tables relating to inferences about the GEV parameters $\mu$, $\sigma$ and $\xi$ are based on the difference between the estimate of a parameter and the estimate based on the corresponding full dataset, for example, $\hat{\mu}_{\text{adjust}} - \hat{\mu}_{\text{full}}$. We present the sample mean, standard deviation and root mean squared error of these differences. Tables concerning 100-block return levels provide statistics to quantify the performances of approaches, absolute and relative to use of the full dataset, in making inferences about the relevant true return level. We present bias, standard deviation and root mean squared error and, owing to the aforementioned skewness of sampling distributions, median bias, inter-quartile range and mean absolute error.

\edit{Table~\ref{tab:gev_table} shows that our adjustment leads to better inferences about the GEV parameters than all other approaches, in the sense of being closer on average to the inferences obtained from the full dataset. For instance, for our approach, the estimates of $\mu$ are generally very close to those from the full dataset.} \edit{The estimates of $\mu$ for the ``weight2'' approach are also reasonably close to the complete data results, but exhibit some positive bias of a generally greater magnitude than the results from our approach. The ``na\"ive'' and ``weight1'' approaches exhibit strong negative bias, as does the ``discard'' approach, albeit to a lesser extent. The estimated standard deviations are very similar for four of the approaches, across all three GEV parameters; the exception is the ``discard'' approach, which generally has much larger standard deviation results.}

\newcolumntype{d}[1]{D{.}{.}{#1}}

\begin{table}[p]
\centering
 \resizebox{0.85\textwidth}{!}{\begin{tabular}{|c|c|d{2.4}d{2.4}d{2.4}|ccc|ccc|} 
 \hline
& & & \multicolumn{1}{c}{bias} & & & sd & & & rmse & \\
distribution & approach & \multicolumn{1}{c}{$\mu$} & \multicolumn{1}{c}{$\sigma$} & \multicolumn{1}{c}{$\xi$} \vline & $\mu$ & $\sigma$ & $\xi$ & $\mu$ & $\sigma$ & $\xi$ \\ \hline \hline
exponential & adjust & -0.0023 & -0.0011 & 0.0004 & 0.065 & 0.058 &
0.058 & 0.065 & 0.058 & 0.058 \\
& weight2 & 0.0034 & 0.0079 & -0.0124 & 0.065 & 0.055 & 0.054 & 0.065 &
0.055 & 0.056 \\
& weight1 & -0.1084 & 0.0028 & -0.0018 & 0.065 & 0.060 & 0.058 & 0.126 &
0.060 & 0.058 \\
& na\"ive & -0.1128 & 0.0026 & -0.0012 & 0.066 & 0.061 & 0.059 & 0.131 &
0.061 & 0.059 \\
& discard & -0.0421 & -0.0197 & -0.0097 & 0.178 & 0.133 & 0.165 & 0.182
& 0.135 & 0.165 \\ \hline
Gaussian & adjust & -0.0003 & -0.0001 & -0.0025 & 0.024 & 0.023 & 0.057
& 0.024 & 0.023 & 0.057 \\
& weight2 & 0.0012 & 0.0025 & -0.0122 & 0.024 & 0.022 & 0.054 & 0.024 &
0.022 & 0.055 \\
& weight1 & -0.0417 & 0.0064 & -0.0046 & 0.024 & 0.024 & 0.057 & 0.048 &
0.025 & 0.057 \\
& na\"ive & -0.0433 & 0.0065 & -0.0042 & 0.025 & 0.024 & 0.058 & 0.050 &
0.025 & 0.058 \\
& discard & -0.0162 & -0.0055 & -0.0149 & 0.068 & 0.052 & 0.158 & 0.070
& 0.052 & 0.159 \\ \hline
Student $t$ & adjust & -0.0041 & 0.0108 & -0.0011 & 0.236 & 0.238 & 0.074
& 0.236 & 0.239 & 0.074 \\
& weight2 & 0.0218 & 0.0359 & -0.0171 & 0.237 & 0.233 & 0.069 & 0.238 &
0.236 & 0.071 \\
& weight1 & -0.3592 & -0.1586 & -0.0026 & 0.231 & 0.239 & 0.075 & 0.427
& 0.287 & 0.075 \\
& na\"ive & -0.3744 & -0.1674 & -0.0019 & 0.234 & 0.240 & 0.075 & 0.441 &
0.293 & 0.075 \\
& discard & -0.1458 & -0.1437 & 0.0074 & 0.613 & 0.583 & 0.202 & 0.630 &
0.600 & 0.202 \\ \hline
beta & adjust & -0.0002 & -0.0002 & 0.0004 & 0.004 & 0.004 & 0.057 &
0.004 & 0.004 & 0.057 \\
& weight2 & 0.0002 & 0.0004 & -0.0121 & 0.004 & 0.003 & 0.053 & 0.004 &
0.003 & 0.055 \\
& weight1 & -0.0070 & 0.0008 & -0.0019 & 0.004 & 0.004 & 0.057 & 0.008 &
0.004 & 0.057 \\
& na\"ive & -0.0073 & 0.0008 & -0.0013 & 0.004 & 0.004 & 0.057 & 0.008 &
0.004 & 0.057 \\
& discard & -0.0026 & -0.0010 & -0.0124 & 0.011 & 0.008 & 0.156 & 0.012
& 0.008 & 0.157 \\ \hline
\end{tabular}}
\caption{Estimation of GEV parameters in comparison to the full data case. The estimated bias, standard deviation (sd) and root mean squared error (rmse) of estimators of $\mu$, $\sigma$ and $\xi$ are given for each approach and for each of the simulation distributions. \edit{The Monte Carlo standard errors associated with these simulation results are provided in Table~\ref{tab:gev_table_mcse} of the Supplementary Material.}}
\label{tab:gev_table}
\end{table}

\begin{table}[!htbp]
\centering
 \resizebox{0.85\textwidth}{!}{\begin{tabular}{|c|c|d{3.3}d{3.3}|d{4.3}d{3.3}|d{5.3}d{3.3}|c|} 
 \hline
& & \multicolumn{1}{c}{bias} & \multicolumn{1}{c}{median bias} \vline & \multicolumn{1}{c}{sd} & \multicolumn{1}{c}{iqr} \vline & \multicolumn{1}{c}{rmse} & \multicolumn{1}{c}{mae} \vline & coverage \\ \hline \hline
exponential & full & 0.025 & -0.119 & 1.190 & 1.496 & 1.190 & 0.913 &
0.950 \\
& adjust & 0.037 & -0.101 & 1.250 & 1.562 & 1.250 & 0.956 & 0.948 \\
& weight2 & -0.056 & -0.187 & 1.188 & 1.518 & 1.189 & 0.928 & 0.948 \\
& weight1 & -0.089 & -0.214 & 1.186 & 1.494 & 1.189 & 0.925 & 0.958 \\
& na\"ive & -0.090 & -0.217 & 1.184 & 1.490 & 1.188 & 0.923 & 0.946 \\
& discard & 0.124 & -0.319 & 2.349 & 2.246 & 2.352 & 1.523 & 0.946 \\ \hline
Gaussian & full & -0.042 & -0.072 & 0.288 & 0.368 & 0.291 & 0.231 &
0.930 \\
& adjust & -0.045 & -0.075 & 0.306 & 0.390 & 0.309 & 0.244 & 0.929 \\
& weight2 & -0.062 & -0.089 & 0.294 & 0.383 & 0.300 & 0.240 & 0.926 \\
& weight1 & -0.073 & -0.102 & 0.293 & 0.372 & 0.302 & 0.241 & 0.941 \\
& na\"ive & -0.073 & -0.102 & 0.292 & 0.373 & 0.301 & 0.241 & 0.925 \\
& discard & -0.045 & -0.135 & 0.636 & 0.551 & 0.638 & 0.380 & 0.929 \\ \hline
Student $t$ & full & 8.924 & -1.535 & 40.896 & 42.116 & 41.856 & 27.735 &
0.965 \\
& adjust & 9.464 & -1.093 & 43.270 & 43.147 & 44.291 & 28.638 & 0.964 \\
& weight2 & 6.217 & -3.436 & 39.597 & 41.110 & 40.080 & 26.949 &
0.962 \\
& weight1 & 4.738 & -4.763 & 39.062 & 39.469 & 39.346 & 26.405 &
0.964 \\
& na\"ive & 4.643 & -4.812 & 38.886 & 39.327 & 39.160 & 26.321 & 0.956 \\
& discard & 29.621 & -3.055 & 312.869 & 62.140 & 314.253 & 55.475 &
0.946 \\ \hline
beta & full & -0.003 & -0.008 & 0.048 & 0.063 & 0.048 & 0.038 & 0.948 \\
& adjust & -0.003 & -0.008 & 0.051 & 0.066 & 0.051 & 0.040 & 0.948 \\
& weight2 & -0.007 & -0.011 & 0.049 & 0.064 & 0.049 & 0.039 & 0.946 \\
& weight1 & -0.008 & -0.012 & 0.049 & 0.063 & 0.049 & 0.039 & 0.957 \\
& na\"ive & -0.008 & -0.012 & 0.049 & 0.063 & 0.049 & 0.039 & 0.945 \\
& discard & -0.003 & -0.019 & 0.091 & 0.094 & 0.091 & 0.063 & 0.941 \\ \hline
\end{tabular}}
\caption{Estimation of the 100-block return level.
The estimated bias, median bias, standard deviation (sd), inter-quartile
range (iqr), root mean squared error (rmse) and mean absolute error
(mae) are given for each approach and for each of the simulation
distributions. The coverage column gives the estimated coverage of
profile-based \(95\%\) confidence intervals. \edit{The Monte Carlo standard errors associated with these simulation results are provided in Table~\ref{tab:rl_table_mcses} of the Supplementary Material.}}
\label{tab:rl_table}
\end{table}

Table~\ref{tab:rl_table} confirms the main findings from Figure~\ref{fig:simulations_RLhistograms}, that our adjustment results in estimated 100-block return levels that are, on average, similar to those based on a full dataset, but are more variable, reflecting the fact that data are missing. The ``na\"ive'' approach tends to result in greater bias than our approach and the discarding of block maxima results in increased variability. \edit{The ``weight1'' approach is again very similar to the na\"ive approach, while the ``weight2'' approach is more successful, but usually more biased than our approach.} In the Student's $t_2$ case, the na\"ive approach outperforms the other approaches in terms of bias but is the poorest when judged using median bias. \edit{The reason for this can be inferred from Figure~\ref{fig:simulations_RLhistograms_Student} of the Supplementary Material. In the Student $t_2$ case, the sampling distributions of the estimators of the 100-year return level are more strongly positively skewed than in the other cases (Figures~\ref{fig:simulations_RLhistograms}, \ref{fig:simulations_RLhistograms_Gaussian} and \ref{fig:simulations_RLhistograms_Beta}), with the effect that the means of these sampling distributions are much greater than their medians. For our adjustment, the median of the sampling distribution is slightly smaller than the true return level, and close to the penultimate approximation to the return level, but its mean is much larger, hence the positive bias. In contrast, the lack of an upwards adjustment when using the ``na\"ive'' approach results in the mean of its sampling distribution being closer to the truth, but its median is much smaller, leading to the relatively large negative median bias.} Our adjustment produces profile-based $95\%$ confidence intervals with estimated coverages that are close to those based on a full dataset, whereas the estimated coverages are lower for the ``na\"ive'' and, to a lesser extent, ``discard'' approaches. The relatively low estimated coverage for the ``na\"ive'' approach is a consequence of its underestimation of statistical uncertainty. \edit{In the exponential and beta cases, the ``full'', ``adjust'' and ``weight2'' approaches have coverages that are closest to the nominal $95\%$, but in the Gaussian and Student's $t_2$ cases,  the estimated coverage for the ``weight1'' and ``na\"ive'' approaches are the closest, respectively.}

\section{Applications}\label{sec:application}
We apply our new methodology to two sets of environmental data. These are both affected by missingness and relate to situations where understanding extremal behaviour may be of interest. The first case study relates to the height of sea surges in Brest, France, while the second concerns ozone levels in Plymouth, U.K. 

\subsection{Case study 1: Brest sea surges}\label{subsec:seasurge}
Sea surges generated during extreme weather events can lead to loss of life and can have enormous economic impacts, a risk exacerbated by recent rises in sea level \citep{Reinert2021}. We analyse sea surge heights measured at high tide at the tide gauge station in Brest, France, between 1846 and 2007, i.e., a total of 162 years. The providers of these data have declustered the raw data to create a series of independent sea surges, \edit{each separated by at least two days}, and applied a correction to account for trends in sea level. \edit{Although this declustering means that the effective block size is smaller than the number of days in a year, the proportion of non-missing raw values in a year should provide a useful measure of the extent to which the corresponding annual maximum is likely to be affected by missingness.}

\begin{figure}[!htbp]
    \centering
    \includegraphics[width=0.75\textwidth]{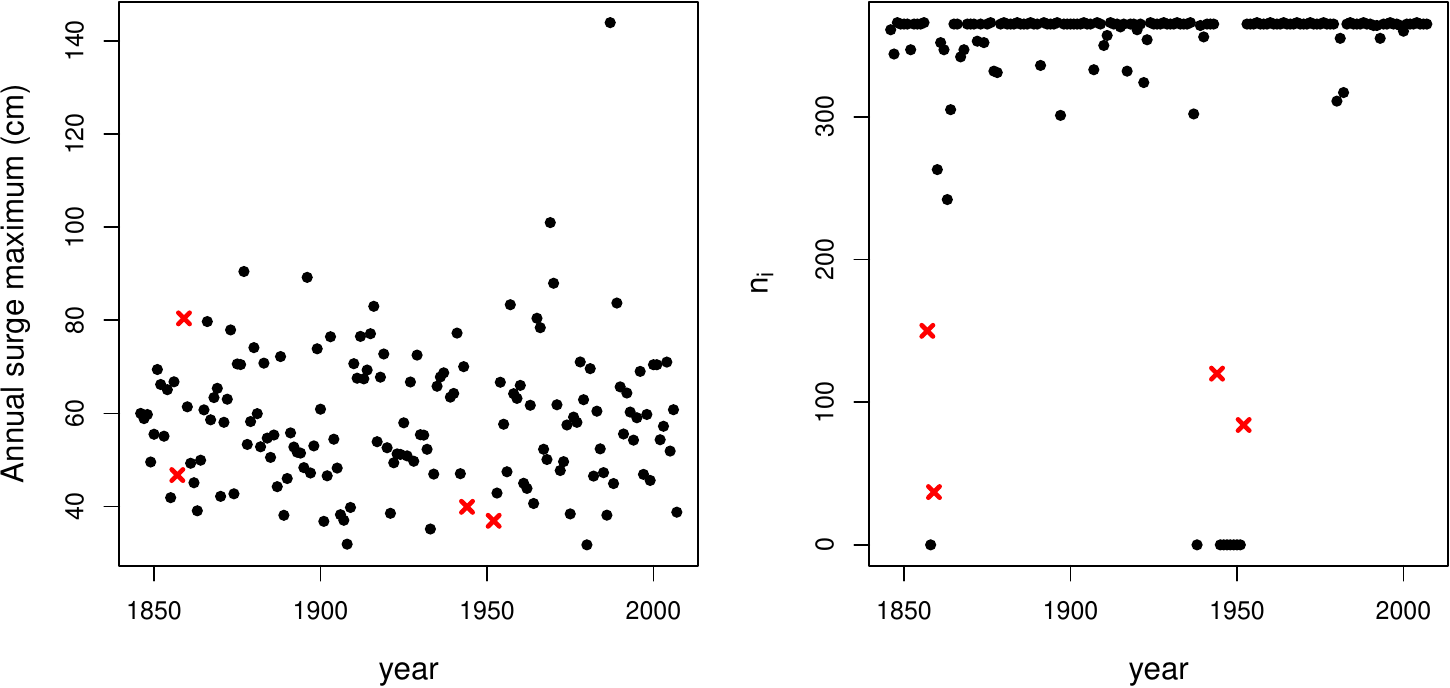}
    \caption{Plots of the maximum recorded sea surge (left) and total number of daily observations (right) per year. The \edit{red crosses} correspond to the four years with the highest proportion of missingness, given that the full year is not missing completely.}
    \label{fig:surgeData}
    \vspace{1cm}
    \includegraphics[width=0.75\textwidth]{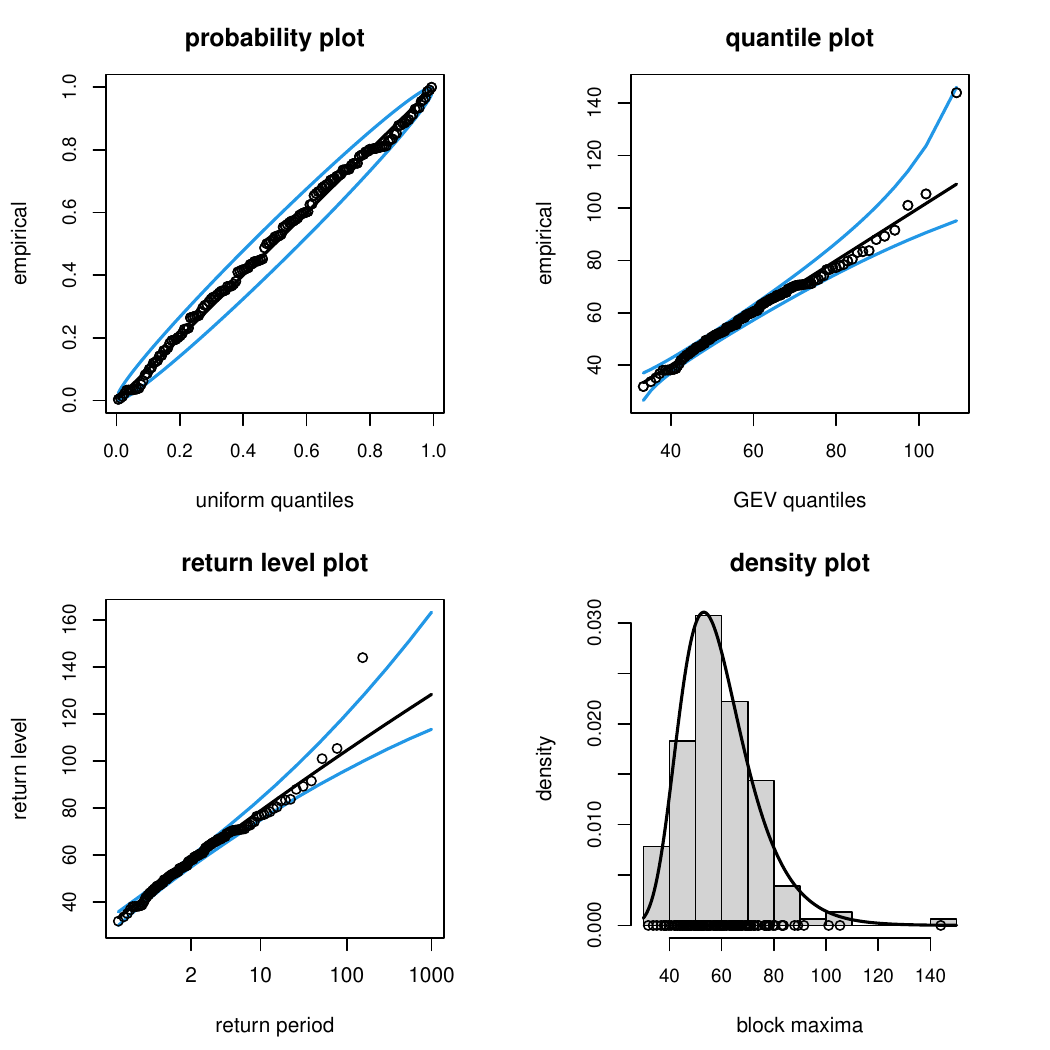}
    \caption{Diagnostic plots for the sea surge data.}
    \label{fig:surgeDiagnostics}
\end{figure}

\begin{table}[!htbp]
\centering
 \resizebox{\textwidth}{!}{\begin{tabular}{|c|c||r r r r|} 
 \hline
\multirow{2}{*}{GEV Model} & Account for missingness?    & \multicolumn{1}{c}{Yes} & \multicolumn{1}{c}{No}  & \multicolumn{1}{c}{Yes} & \multicolumn{1}{c}{No} \vline \\
      & Remove 1857, 1859, 1944, 1952?       & \multicolumn{1}{c}{No} & \multicolumn{1}{c}{No} & \multicolumn{1}{c}{Yes} & \multicolumn{1}{c}{Yes} \vline \\ [0.5ex] 
 \hline\hline
\multirow{3}{*}{Parameter} & $\mu$      & 52.89 (1.07)  & 52.27 (1.07)    & 52.84 (1.08)     & 52.57 (1.08) \\ 
                           & $\sigma$   & 11.84 (0.74)   & 12.09 (0.76)     & 11.93 (0.75)      & 12.01 (0.76) \\
                           & $\xi$      & $-$0.02 (0.04)    & $-$0.03 (0.04)     & $-$0.03 (0.04)       &  $-$0.03 (0.04) \\
\hline
\multirow{3}{*}{Return period} & $25$ & 89.4 (84.0,97.6) & 89.1 (83.9,97.1) & 89.2 (83.9,97.2) & 89.1 (83.8,97.0) \\
& 50 & 97.0 (90.3,108.5) & 96.8 (90.2,107.9) & 96.8 (90.2,107.8) & 96.6 (90.1,107.5) \\
& 100 & 104.5 (96.3,120.1) & 104.2 (96.2,119.3) & 104.1 (96.1,119.0) & 103.9 (96.0,118.6) \\ \hline
\end{tabular}}
 \caption{\edit{Rows 1-3: GEV parameter estimates for the Brest sea surge data with four different modelling choices. Numbers in parentheses represent the standard errors of the parameter estimates. Rows 4-6: return level estimates. Numbers in parentheses are profile-based 95\% confidence intervals. The first column of results relates to our proposed method.}}
 \label{tab:surgeEstimates}
 \end{table}
 
Figure~\ref{fig:surgeData} shows the maximum recorded sea surge in each year and the respective number of non-missing daily observations. Many (113) years do not have any missing data, but overall, approximately 9\% of the raw data are missing. There are nine years during which no data were recorded, including the years 1945--51 during and following World War II. For the years 1857, 1859, 1944 and 1952 more than 50\% of the daily values were missing. For three of these years, the annual maximum is relatively low, but not unusually so. The diagnostic plots in Figure~\ref{fig:surgeDiagnostics} relate to our new model. Overall, the fit of the model is good after accounting for missingness, although the largest observation lies above the upper limit of its confidence interval 95\% in the return level plot.

Table~\ref{tab:surgeEstimates} provides a comparison between the inferences using our adjustment (column three) and the na\"ive approach of ignoring missingness (column four). For these data, the differences are not substantial, but they are consistent with the estimated biases in Section~\ref{sec:simulations}. In particular, for the na\"ive approach, the estimate of $\mu$ (52.27cm) is smaller than that from our approach (52.89cm). \edit{The estimated return levels are also slightly smaller using the na\"ive approach than after making our adjustment.} Column six of Table~\ref{tab:surgeEstimates} gives results for the ``discard'' approach, i.e., discarding the four years with more than 50\% of missing daily values and making no other adjustment. As we expect, this reduces the amount by which the estimate of $\mu$ has decreased relative to our adjustment. The results in column five are produced by the strategy of removing these four years of data {\bf and} making our adjustment. If our adjustment provides a sensible adjustment even when some block maxima have high levels of missingness, then we expect the results in columns three and five to be similar, with slightly increased standard errors for the latter. This is what we observe.

\subsection{Case study 2: Plymouth ozone levels}\label{subsec:ozone}
\begin{figure}[!htbp]
    \centering
    \includegraphics[width=0.75\textwidth]{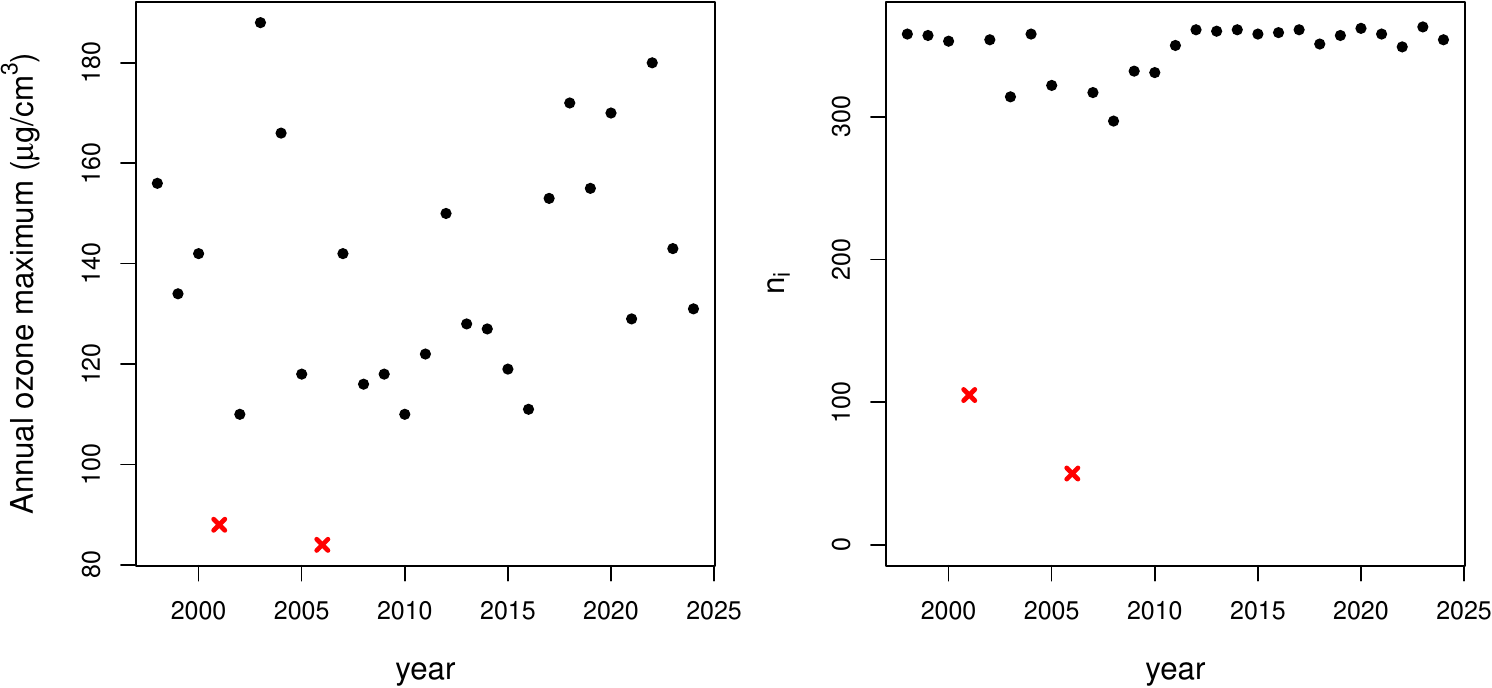}
    \caption{Plots of the maximum recorded ozone level (left) and total number of daily observations (right) per year. Observations corresponding to the two years with the highest proportion of missingness are shown by the \edit{red crosses}.}
    \label{fig:ozoneData}
    \vspace{1cm}
    \includegraphics[width=0.75\textwidth]{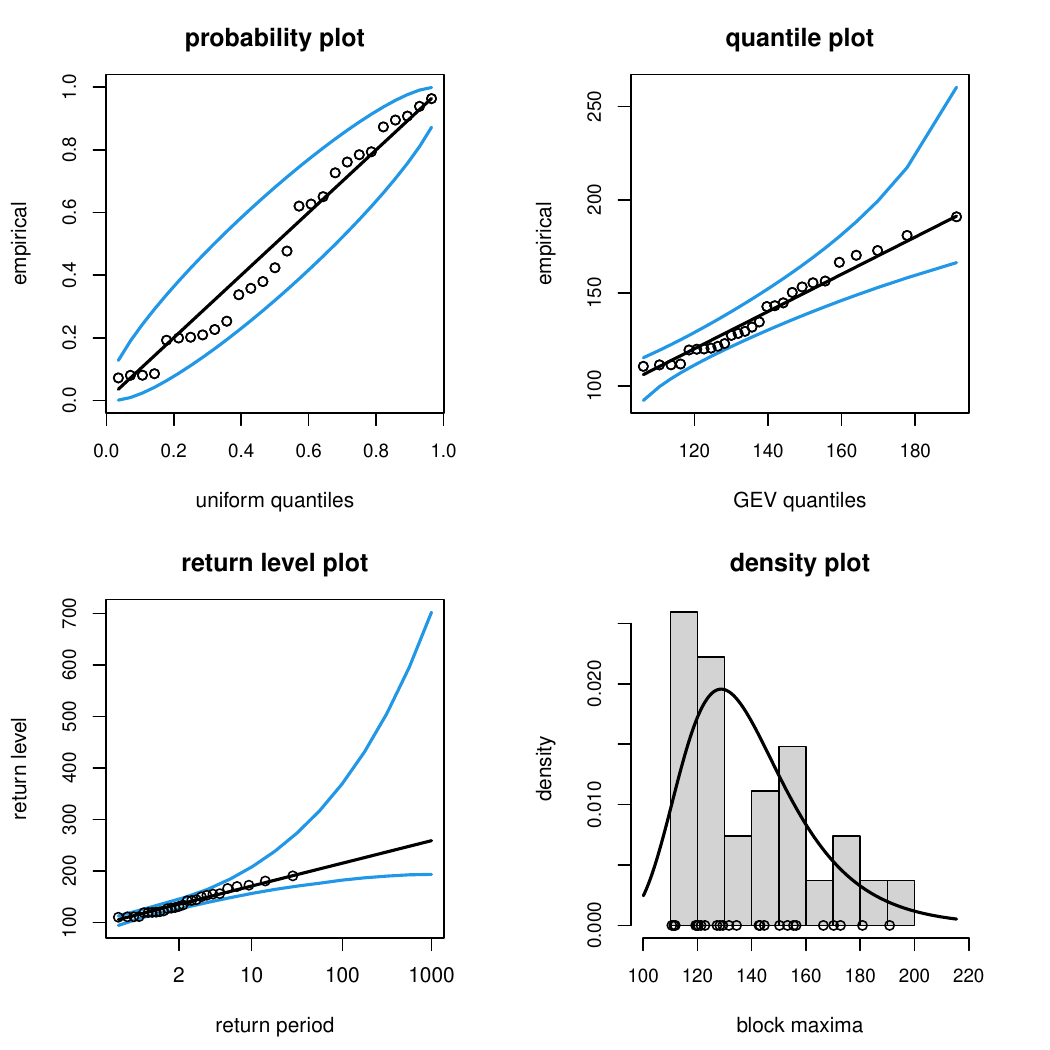}
    \caption{Diagnostic plots for the ozone data.}
    \label{fig:ozoneDiagnostics}
\end{figure}

It is well known that high levels of air pollution can be detrimental to human health. One widely studied component of air pollution is ozone (O$_3$), which has links to conditions such as chronic respiratory disease \citep[see, e.g.,][]{Malashock2022}. A recent study suggests that global health-related risks of ozone may have previously been underestimated \citep{Wang2025}. This emphasises the importance of monitoring ozone levels, with estimation of future extremes potentially aiding mitigation efforts and instructing policy implementation.

In this second case study, we consider ozone levels measured in Plymouth, U.K., between 1998 and 2024, i.e., a total of 27 years. The raw data are measured in micrograms per metre cubed ($\mu$g$/$m$^3$), with daily maximum observations recorded. Overall, approximately 10\% of the raw data are missing, with some variation in the proportion of missingness per year. In Figure~\ref{fig:ozoneData}, we show the maximum recorded ozone value in each year, as well as the respective number of non-missing daily observations. The diagnostic plots in Figure~\ref{fig:ozoneDiagnostics} relate to our new model. In this case, the density histogram is not the most useful due to the small number of block maximum observations available, but otherwise the plots confirm that a good fit has been achieved when we account for missingness. 

\edit{We provide estimates of the GEV parameters and selected return levels} from our modelling approach in Table~\ref{tab:ozoneEstimates}, alongside results for three other sets of modelling choices. As in the simulation study, one option is to completely ignore the missingness in the raw data and fit a standard GEV model to the observed annual maxima. In this case, we see quite different parameter estimates, particularly for the shape parameter, taking the estimated GEV distribution from one with a light, unbounded upper tail ($\xi=0$) to one with a finite upper bound ($\xi<0$) and highlighting that failure to account for missingness can lead to spurious results. \edit{Similarly, the point estimate of the 100-year return level using the na\"ive approach is 192$\mu$g$/$m$^3$, which is substantially lower than the estimates from the other approaches. The upper limit of the 95\% confidence interval for the 100-year return level is much lower than those of the other approaches, a result of the strongly negative estimate $\hat{\xi} = -0.28$. }

\begin{table}[t]
\centering
 \resizebox{\textwidth}{!}{\begin{tabular}{|c|c||r r r r|} 
 \hline
\multirow{2}{*}{GEV Model} & Account for missingness?    & \multicolumn{1}{c}{Yes} & \multicolumn{1}{c}{No}  & \multicolumn{1}{c}{Yes} & \multicolumn{1}{c}{No} \vline \\
      & Remove 2001 and 2006?       & \multicolumn{1}{c}{No} & \multicolumn{1}{c}{No} & \multicolumn{1}{c}{Yes} & \multicolumn{1}{c}{Yes} \vline \\ [0.5ex] 
 \hline\hline
\multirow{3}{*}{Parameter} & $\mu$      & 128.77 (4.40)  & 126.52 (5.53)    & 129.55 (4.60)     & 128.59 (4.55) \\ 
                           & $\sigma$   & 18.81 (2.63)   & 25.50 (4.00)     & 17.65 (3.41)      & 17.78 (3.61) \\
                           & $\xi$      & 0.00 (0.16)    & $-$0.28 (0.15)     & 0.04 (0.26)       &  0.04 (0.27) \\
\hline
\multirow{3}{*}{Return period} & $25$ & 189 (168,259) & 180 (169,215) & 190 (168,330) & 189 (168,324) \\
& 50 & 202 (175,309) & 187 (174,235) & 204 (175,461) & 203 (175,454) \\
& 100 & 216 (181,370) & 192 (179,255) & 219 (180,668) & 217 (180,658) \\ \hline
\end{tabular}}
 \caption{\edit{Rows 1-3: GEV parameter estimates for the Brest sea surge data with four different modelling choices. Numbers in parentheses represent the standard errors of the parameter estimates. Rows 4-6: return level estimates. Numbers in parentheses are profile-based 95\% confidence intervals. The first column of results relates to our proposed method.}}
 \label{tab:ozoneEstimates}
 \end{table}

For this dataset, there are two years with far fewer observations than the rest: 2001 with 105 daily recordings, and 2006 with just 50. Understandably, these are also the years with the lowest observed block maxima values, as highlighted in Figure~\ref{fig:ozoneData}. As in the previous case study, it is natural to investigate the effect of removing these observations from our analysis. The parameter estimates for this reduced dataset, for both our new approach and the standard GEV model, are also shown in Table~\ref{tab:ozoneEstimates}. In both cases, the point estimates are much closer to those originally obtained for our proposed method, although removing some observations has induced \edit{larger standard errors, and resulted in much larger upper confidence limits for the return levels.}. In particular, these results support the conclusion that $\xi$ is close to zero, with the distribution having no finite upper end point. 

\begin{figure}
    \centering
    \includegraphics[width=0.5\textwidth]{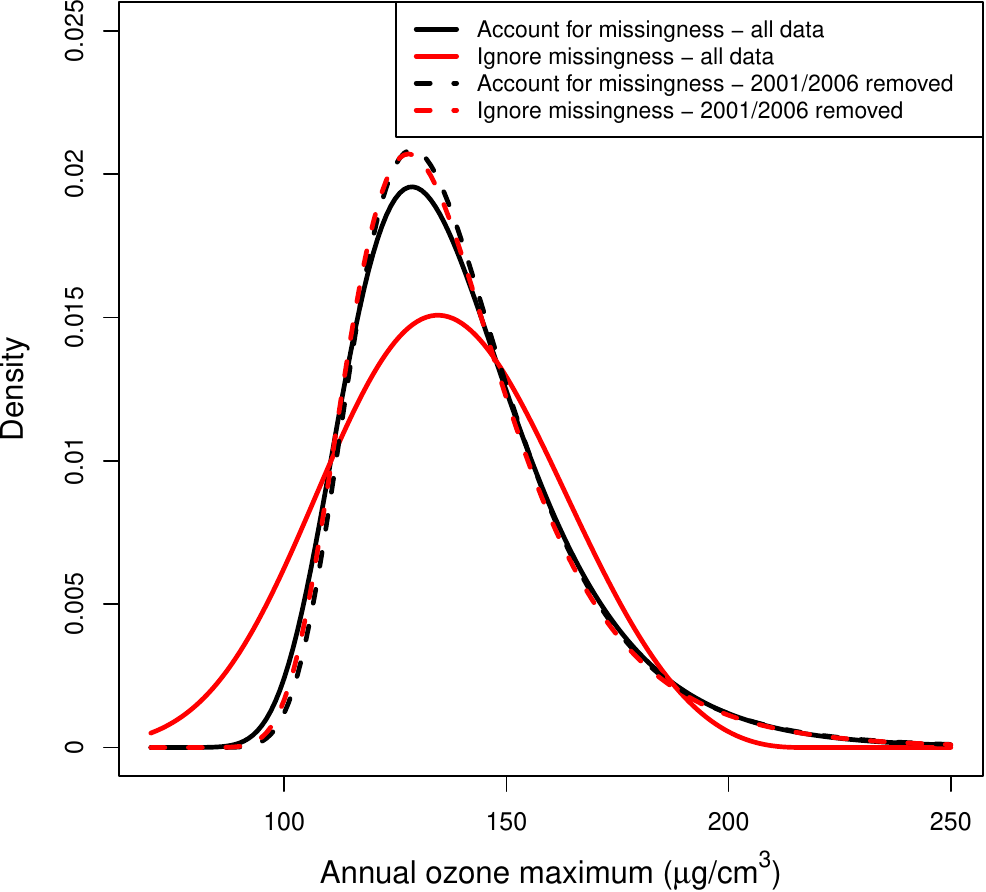}
    \caption{Comparison of GEV densities for the four sets of parameter estimates in Table~\ref{tab:ozoneEstimates}.}
    \label{fig:ozoneDensities}
\end{figure}

To further facilitate comparison, we plot the estimated GEV densities for all four approaches in Figure~\ref{fig:ozoneDensities}. We observe that if we take the standard GEV modelling approach and include all observed block maxima, the two smallest observations (which are unlikely to accurately represent the true annual maxima due to missingness) significantly impact the results. As expected from the parameter estimates, removing the years 2001 and 2006 yields similar estimated GEV densities to our proposed method in both cases; this is particularly apparent in the upper tails of the distributions, indicating that these would lead to very similar return level estimates. This emphasises the reliability of our approach over the standard GEV model applied to all 27 years, since for a given model, we should ideally see stable parameter estimates when removing some observations. It also underlines the potential pitfalls of ignoring missingness completely when modelling block maxima. \edit{To appreciate why the two lowest block maxima values in Figure~\ref{fig:ozoneData} have such an impact on the estimates of the GEV parameters and return levels, it is useful to consider the influence function \citep{Hampel2005}, which measures the effect on a parameter estimator of changing one observation in a sample; further discussion on this can be found in Section~\ref{sm:influence} of the Supplementary Material.} 

Although removing the problematic observations and fitting a GEV distribution as usual gave similar results to our proposed approach, we have the benefit that no threshold for acceptable missingness needs to be chosen, and we do not need to rely on so many of the years having close to a full complement of raw data to obtain reliable results. Given that only 27 years' worth of data are available to begin with, it is also preferable to retain as many observations as possible to avoid unnecessarily inflating the uncertainty in our estimates.

\section{Discussion}\label{sec:discussion}
The aim of this paper is to introduce a first approach for handling missing data when modelling block maxima. We present a simple yet effective model that builds on existing theory, adapting the GEV distribution to allow for blocks with varying numbers of observations. Our model has the benefit of parsimony, since we require exactly the same three parameters that are used in the standard block maxima approach, but despite its simplicity is competitive even with the ideal scenario where all data are observed.

We see our contribution as one of the first steps in the handling of missing data in extreme value contexts, and acknowledge that there are various extensions that could be made. In particular, we work under assumptions that are likely to be too strict in some practical applications, namely that the underlying data are independent and identically distributed and that the missingness is non-informative. There are some cases where violations of these assumptions may not impact the results, but there are others where further work is required to properly model the missingness. We discuss some possible extensions below.

One way in which the assumption of identically distributed data can be violated is through seasonality, which is commonly seen in environmental applications. In block maxima modelling, this issue is generally dealt with by taking blocks that are equal to one year in length (or multiples of one year if required for the asymptotic results to be justified). Even if the underlying data exhibit seasonality, the GEV distribution will often provide a good model for the block maxima \citep{Coles2001}. In our case, the same is true as long as the level of missingness is approximately stationary throughout the year. If the proportion of missingness varies through the year, e.g., if seasons with generally higher observations also have higher levels of missingness, the model in~\eqref{eqn:locscale_miss} will be misspecified, leading again to biased parameter and return level estimates. An option here, commonly used in the extreme value literature, is to concentrate the analysis only on the times of year where extremes are most likely to occur, leading to approximate stationarity in the raw data while still providing useful inference. \edit{As seen in our simulation study, reliable inference can be achieved using our approach for blocks corresponding to single seasons (with $n=90$).} For other types of non-stationarity, e.g., overall temporal trends in the data, is it common to include covariates within the GEV model parameters; with non-informative missingness, this same technique could be used within our modelling framework.

The assumption of independence is often unrealistic, with data exhibiting short-term temporal dependence, leading to clusters of extremes. If time series data follow a stationary sequence satisfying a condition that restricts the long-term impact of dependence on extremes, and with a marginal distribution for which the extremal types theorem in Section~\ref{sec:ETT} applies, then a limiting GEV distribution still arises for block maxima of these dependent data \citep{Leadbetter1983}. Therefore, the limiting GEV distribution is used routinely as a model for block maxima of stationary time series even when short-term dependence is expected. If missingness is non-informative then we expect the benefits of our adjustment to be realised in this more general setting. \edit{We provide a small simulation study in Section~\ref{sm:non-iid} of the Supplementary Material to investigate this point further. Indeed, while our model sees some bias introduced through failure to properly account for the temporal dependence, it does still outperform the alternative approaches. Further investigating this issue, and suitably adapting our approach, is left to future work.}

It is also quite likely for informative missingness to occur in environmental applications, but this is not accounted for in our current approach. It is easy to imagine situations where the most extreme events are the ones that are the hardest to reliably record, e.g., river flow gauges being damaged by fast-running water, and there is a need to be able to capture such phenomena. To account for this, one could attempt to explicitly model the missingness mechanism and incorporate this into the GEV model. One option is to take a regression-based approach, exploiting covariates that help to explain whether or not values are missing.

We have focused our attention on one classical model for univariate extremes, but similar issues can arise when applying other models for block maxima, such as the blended generalised extreme value (bGEV) distribution of \cite{CastroCamilo2022}, and should also be considered in other extreme value contexts. There are many extreme value methods that extend beyond block maxima modelling, and another natural area for further work is in the development of methods to handle missing data in these different frameworks. \edit{The standard alternative to block maxima modelling for univariate extremes is to model threshold exceedances using a generalised Pareto distribution (GPD).} We hypothesise that for non-informative missingness, there is less of a problem for modelling threshold exceedances than block maxima, since the usual GPD assumption will continue to marginally hold. A more interesting question comes from studying the inter-exceedance times under temporal dependence, which will be censored under missingness, making estimation a challenge. \edit{Another, less commonly used, extension of the block maxima framework is to model the $r$-largest observations in each block through a generalisation of the GEV model (see \citet[][Section~3.5]{Coles2001}). Ignoring missingness in this setting would have similar consequences to those discussed in this paper, with biased parameter estimates and underestimated return levels. However, it is not immediately clear how our modelling strategy extends to this setting, and this presents a potentially interesting challenge for further research.}

In the context of spatial extremes, the challenge of missing data is considered by \cite{Healy2025}, who also point out the general lack of missing data considerations in the extreme value analysis literature. In their discussion contribution, \cite{Richards2025_discussion} follow up on this point, investigating how different missingness mechanisms can differently impact results for $r$-Pareto processes specifically. Their findings reiterate the need for careful consideration of these missing data issues and the development of related methodology.

\bibliographystyle{apalike}
\bibliography{refs}

\newpage

\renewcommand{\thesection}{\Alph{section}}
\renewcommand{\thefigure}{\Alph{figure}}
\renewcommand{\thetable}{\Alph{table}}
\renewcommand{\theequation}{\Alph{equation}}
\setcounter{section}{0}
\setcounter{figure}{0}
\setcounter{table}{0}
\setcounter{equation}{0}

\begin{center}
\hrule\vspace{10pt}\hrule\vspace{10pt} 
{\Large Supplementary Material for\\ ``Accounting for missing data when modelling block maxima''}\\
Emma S.\ Simpson and Paul J.\ Northrop\\
\vspace{20pt}\hrule\vspace{10pt}\hrule
\end{center}

\section{Missingness-dependent GEV parameters}\label{sm:GEVmissingParameters}

Here, we show how to obtain the missingness-dependent location and scale parameters for our proposed GEV model, as stated in Section~\ref{subsec:parsimony} of the main paper. For $\xi\neq 0$,
\begin{align}
    \left\{G(z;\mu,\sigma,\xi)\right\}^{n_i/n} &= \exp\left[-\frac{n_i}{n}\left\{1+\xi\left(\frac{z-\mu}{\sigma}\right)\right\}_+^{-1/\xi}\right]\nonumber\\
    &=\exp\left[-\left\{\frac{\xi}{\sigma \left(n_i/n\right)^\xi}\left(\frac{\sigma}{\xi} + z-\mu\right)\right\}_+^{-1/\xi}\right]\nonumber\\
    &=\exp\left[-\left\{1 + \frac{\xi}{\sigma \left(n_i/n\right)^\xi}\left(\frac{\sigma}{\xi} + z-\mu-\frac{\sigma \left(n_i/n\right)^\xi}{\xi}\right)\right\}_+^{-1/\xi}\right]\nonumber\\
    &=\exp\left[-\left\{1 + \frac{\xi}{\sigma \left(n_i/n\right)^\xi}\left(z-\left[\mu+\frac{\sigma}{\xi}\left\{ \left(\frac{n_i}{n}\right)^\xi-1\right\}\right]\right)\right\}_+^{-1/\xi}\right]\nonumber\\
    &=\exp\left[-\left\{1+\xi\left(\frac{z-\mu(n_i)}{\sigma(n_i)}\right)\right\}_+^{-1/\xi}\right],
    \label{eqn:GEVparam_derivation}
\end{align}
with 
\begin{align}
    \mu(n_i) &= \mu+\frac{\sigma}{\xi}\left\{\left(\frac{n_i}{n}\right)^\xi-1\right\},\nonumber\\
    \sigma(n_i) &= \sigma \left(\frac{n_i}{n}\right)^\xi.
    \label{eqn:GEVparam_derivation2}
\end{align}
For $\xi=0$, as is standard for the GEV distribution, an analogous result holds by considering the limits of~\eqref{eqn:GEVparam_derivation}~and~\eqref{eqn:GEVparam_derivation2} as $\xi\rightarrow 0$. This yields the usual form of the GEV distribution for $\xi=0$, with $\mu(n_i) = \mu + \sigma\log(n_i/n)$ and $\sigma(n_i) = \sigma$.

\section{Versions of Figure \ref{fig:simulations_RLhistograms} for other distributions} \label{sm:figure3} \vspace{0.25cm}

\begin{figure}[!htbp]
    \centering
    \includegraphics[width=0.9\textwidth]{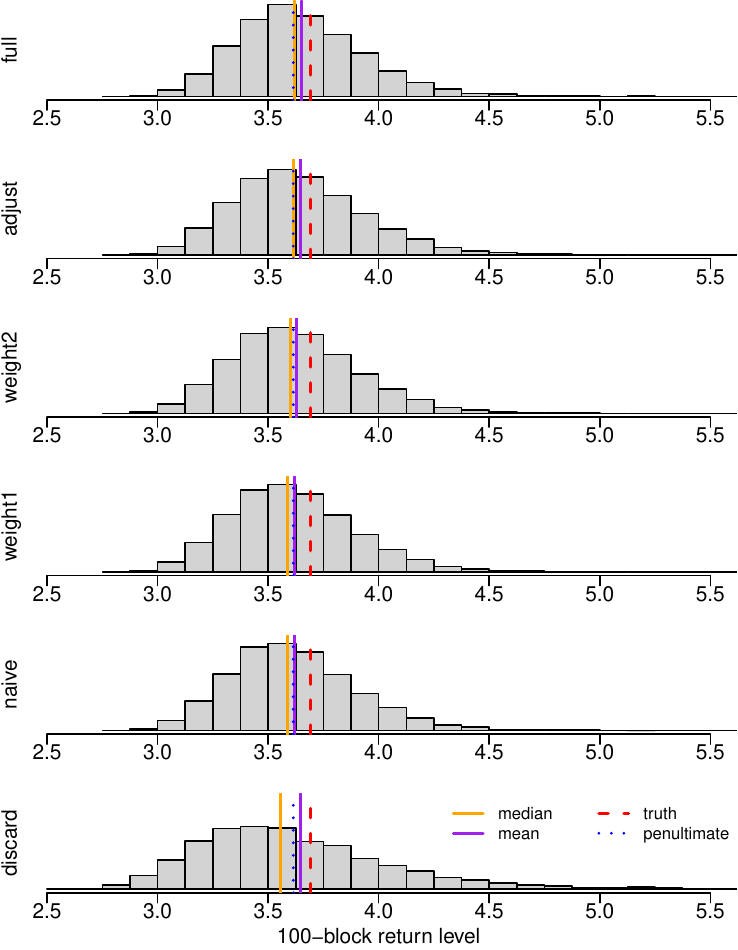}
    \caption{Histograms of estimated 100-block return levels across 10,000 simulation replicates for a standard Gaussian distribution. }
    \label{fig:simulations_RLhistograms_Gaussian}
\end{figure}

\begin{figure}[p]
    \centering
    \includegraphics[width=0.9\textwidth]{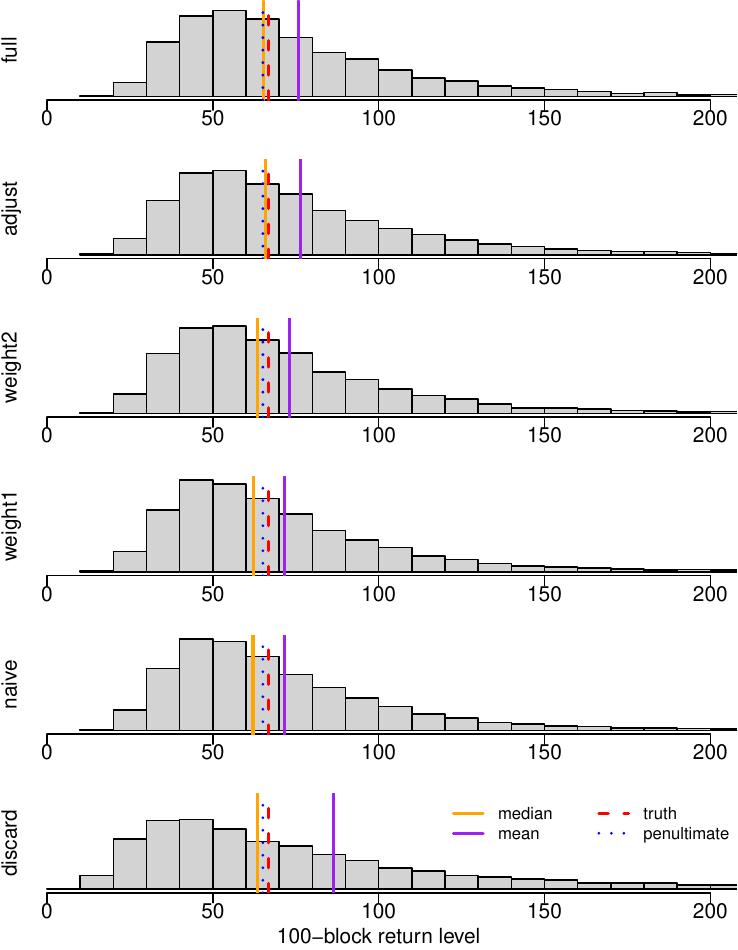}
    \caption{Histograms of estimated 100-block return levels across 10,000 simulation replicates for a Student's $t_2$ distribution. }
    \label{fig:simulations_RLhistograms_Student}
\end{figure}

\begin{figure}[p]
    \centering
    \includegraphics[width=0.9\textwidth]{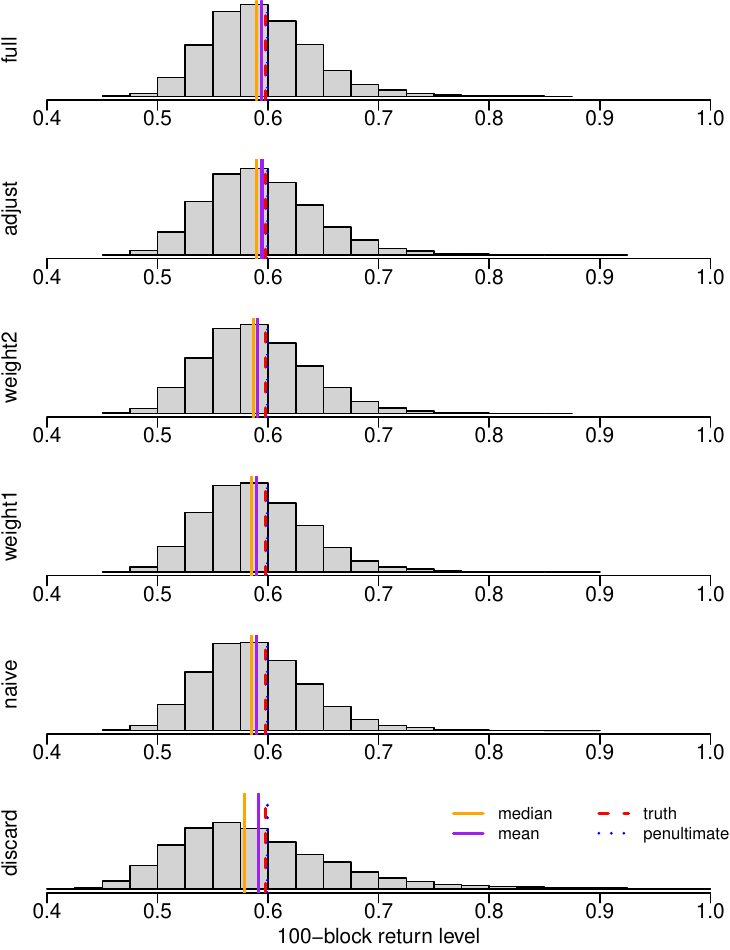}
    \caption{Histograms of estimated 100-block return levels across 10,000 simulation replicates for a Beta(1, 10) distribution. }
    \label{fig:simulations_RLhistograms_Beta}
\end{figure}

\section{Monte Carlo standard errors for Tables~\ref{tab:gev_table}~and~\ref{tab:rl_table}}\label{sm:mcse}

In this section, we present tables of the Monte Carlo standard errors for the simulation results presented in Section~\ref{subsec:SSresults} of the main paper. Table~\ref{tab:gev_table_mcse} relates to the estimates for the individual GEV parameters, while Table~\ref{tab:rl_table_mcses} corresponds to the 100-block return level estimates.\\
 
\begin{table}[!htbp]
\centering
 \resizebox{\textwidth}{!}{
 \begin{tabular}{|c|c|ccc|ccc|ccc|} 
 \hline
& & & \multicolumn{1}{c}{bias} & & & sd & & & rmse & \\
distribution & approach & \multicolumn{1}{c}{$\mu$} & \multicolumn{1}{c}{$\sigma$} & \multicolumn{1}{c}{$\xi$} \vline & $\mu$ & $\sigma$ & $\xi$ & $\mu$ & $\sigma$ & $\xi$ \\ \hline \hline
exponential & adjust & 0.00065 & 0.00058 & 0.00058 & 0.00053 & 0.00056 &
0.00069 & 0.00054 & 0.00056 & 0.00069 \\
& weight2 & 0.00065 & 0.00055 & 0.00054 & 0.00052 & 0.00048 & 0.00069 &
0.00051 & 0.00050 & 0.00074 \\
& weight1 & 0.00065 & 0.00060 & 0.00058 & 0.00054 & 0.00057 & 0.00069 &
0.00071 & 0.00058 & 0.00070 \\
& na\"ive & 0.00066 & 0.00061 & 0.00059 & 0.00055 & 0.00058 & 0.00070 &
0.00073 & 0.00058 & 0.00070 \\
& discard & 0.00178 & 0.00133 & 0.00165 & 0.00142 & 0.00109 & 0.00174 &
0.00139 & 0.00110 & 0.00175 \\ \hline
Gaussian & adjust & 0.00024 & 0.00023 & 0.00057 & 0.00019 & 0.00022 &
0.00074 & 0.00019 & 0.00022 & 0.00076 \\
& weight2 & 0.00024 & 0.00022 & 0.00054 & 0.00018 & 0.00019 & 0.00075 &
0.00018 & 0.00019 & 0.00081 \\
& weight1 & 0.00024 & 0.00024 & 0.00057 & 0.00019 & 0.00022 & 0.00074 &
0.00027 & 0.00025 & 0.00076 \\
& na\"ive & 0.00025 & 0.00024 & 0.00058 & 0.00020 & 0.00023 & 0.00074 &
0.00027 & 0.00026 & 0.00076 \\
& discard & 0.00068 & 0.00052 & 0.00159 & 0.00054 & 0.00043 & 0.00172 &
0.00053 & 0.00043 & 0.00172 \\ \hline
Student $t$ & adjust & 0.00236 & 0.00238 & 0.00074 & 0.00202 & 0.00222 &
0.00078 & 0.00203 & 0.00223 & 0.00079 \\
& weight2 & 0.00237 & 0.00233 & 0.00069 & 0.00193 & 0.00205 & 0.00072 &
0.00189 & 0.00207 & 0.00080 \\
& weight1 & 0.00231 & 0.00239 & 0.00075 & 0.00205 & 0.00221 & 0.00078 &
0.00261 & 0.00223 & 0.00079 \\
& na\"ive & 0.00234 & 0.00240 & 0.00075 & 0.00209 & 0.00222 & 0.00079 &
0.00265 & 0.00225 & 0.00079 \\
& discard & 0.00613 & 0.00583 & 0.00202 & 0.00544 & 0.00528 & 0.00205 &
0.00512 & 0.00506 & 0.00205 \\ \hline
beta & adjust & 0.00004 & 0.00004 & 0.00057 & 0.00003 & 0.00003 &
0.00066 & 0.00003 & 0.00003 & 0.00066 \\
& weight2 & 0.00004 & 0.00003 & 0.00053 & 0.00003 & 0.00003 & 0.00066 &
0.00003 & 0.00003 & 0.00071 \\
& weight1 & 0.00004 & 0.00004 & 0.00057 & 0.00003 & 0.00003 & 0.00066 &
0.00005 & 0.00004 & 0.00067 \\
& na\"ive & 0.00004 & 0.00004 & 0.00057 & 0.00003 & 0.00003 & 0.00066 &
0.00005 & 0.00004 & 0.00067 \\
& discard & 0.00011 & 0.00008 & 0.00157 & 0.00009 & 0.00007 & 0.00154 &
0.00009 & 0.00007 & 0.00154 \\ \hline
\end{tabular}}
\caption{Monte Carlo standard errors associated with Table~\ref{tab:gev_table} of the main paper.}
\label{tab:gev_table_mcse}
\end{table}

\newcolumntype{d}[1]{D{.}{.}{#1}}

\begin{table}[!htbp]
\centering
 \resizebox{\textwidth}{!}{\begin{tabular}{|c|c|d{2.4}d{2.4}|d{3.5}d{2.5}|d{4.5}d{2.5}|c|} 
 \hline
& & \multicolumn{1}{c}{bias} & \multicolumn{1}{c}{median bias} \vline & \multicolumn{1}{c}{sd} & \multicolumn{1}{c}{iqr} \vline & \multicolumn{1}{c}{rmse} & \multicolumn{1}{c}{mae} \vline & coverage \\ \hline \hline
exponential & full & 0.0119 & 0.0138 & 0.0128 & 0.0186 & 0.0129 & 0.0076
& 0.0022 \\
& adjust & 0.0125 & 0.0159 & 0.0136 & 0.0185 & 0.0137 & 0.0080 &
0.0022 \\
& weight2 & 0.0119 & 0.0166 & 0.0118 & 0.0186 & 0.0115 & 0.0075 &
0.0022 \\
& weight1 & 0.0119 & 0.0133 & 0.0125 & 0.0179 & 0.0121 & 0.0074 &
0.0020 \\
& na\"ive & 0.0118 & 0.0142 & 0.0125 & 0.0179 & 0.0120 & 0.0074 &
0.0023 \\
& discard & 0.0235 & 0.0223 & 0.0925 & 0.0301 & 0.0930 & 0.0180 &
0.0023 \\ \hline
Gaussian & full & 0.0029 & 0.0031 & 0.0027 & 0.0052 & 0.0026 & 0.0018 &
0.0025 \\
& adjust & 0.0031 & 0.0037 & 0.0033 & 0.0048 & 0.0031 & 0.0019 &
0.0026 \\
& weight2 & 0.0029 & 0.0036 & 0.0026 & 0.0050 & 0.0024 & 0.0018 &
0.0026 \\
& weight1 & 0.0029 & 0.0031 & 0.0031 & 0.0041 & 0.0028 & 0.0018 &
0.0024 \\
& na\"ive & 0.0029 & 0.0033 & 0.0031 & 0.0049 & 0.0028 & 0.0018 &
0.0026 \\
& discard & 0.0064 & 0.0049 & 0.0713 & 0.0064 & 0.0708 & 0.0051 &
0.0026 \\ \hline
Student $t$ & full & 0.4090 & 0.3383 & 0.7072 & 0.5775 & 0.7522 & 0.3156 &
0.0018 \\
& adjust & 0.4327 & 0.4659 & 0.9207 & 0.5687 & 0.9620 & 0.3408 &
0.0019 \\
& weight2 & 0.3960 & 0.3525 & 0.7094 & 0.5716 & 0.7428 & 0.2969 &
0.0019 \\
& weight1 & 0.3906 & 0.3757 & 0.7740 & 0.4624 & 0.7998 & 0.2943 &
0.0019 \\
& na\"ive & 0.3889 & 0.4169 & 0.7682 & 0.5288 & 0.7936 & 0.2872 &
0.0020 \\
& discard & 3.1287 & 0.6265 & 120.2303 & 1.0477 & 119.9501 & 3.1176 &
0.0023 \\ \hline
beta & full & 0.0005 & 0.0005 & 0.0004 & 0.0007 & 0.0004 & 0.0003 &
0.0022 \\
& adjust & 0.0005 & 0.0005 & 0.0005 & 0.0007 & 0.0005 & 0.0003 &
0.0022 \\
& weight2 & 0.0005 & 0.0006 & 0.0004 & 0.0008 & 0.0004 & 0.0003 &
0.0023 \\
& weight1 & 0.0005 & 0.0005 & 0.0004 & 0.0007 & 0.0004 & 0.0003 &
0.0020 \\
& na\"ive & 0.0005 & 0.0006 & 0.0004 & 0.0008 & 0.0004 & 0.0003 &
0.0023 \\
& discard & 0.0009 & 0.0008 & 0.0029 & 0.0013 & 0.0028 & 0.0007 &
0.0024 \\ \hline
\end{tabular}}
\caption{Monte Carlo standard errors associated with Table~\ref{tab:rl_table} of the main paper.}
\label{tab:rl_table_mcses}
\end{table}

\section{Influence functions for the Plymouth ozone data}\label{sm:influence}

In Section~\ref{subsec:ozone} of the main paper, we demonstrated that for the Plymouth ozone data, there were two block maxima observations that appeared to have the largest effect on the estimation of our GEV model. These corresponded to the two years most affected by missingness. Here, we investigate this point further by considering the influence function \citep{Hampel2005} of the GEV parameters and return levels.

\begin{figure}[!ht]
    \begin{subfigure}{0.49\textwidth}
    \centering
    \includegraphics[height=7cm]{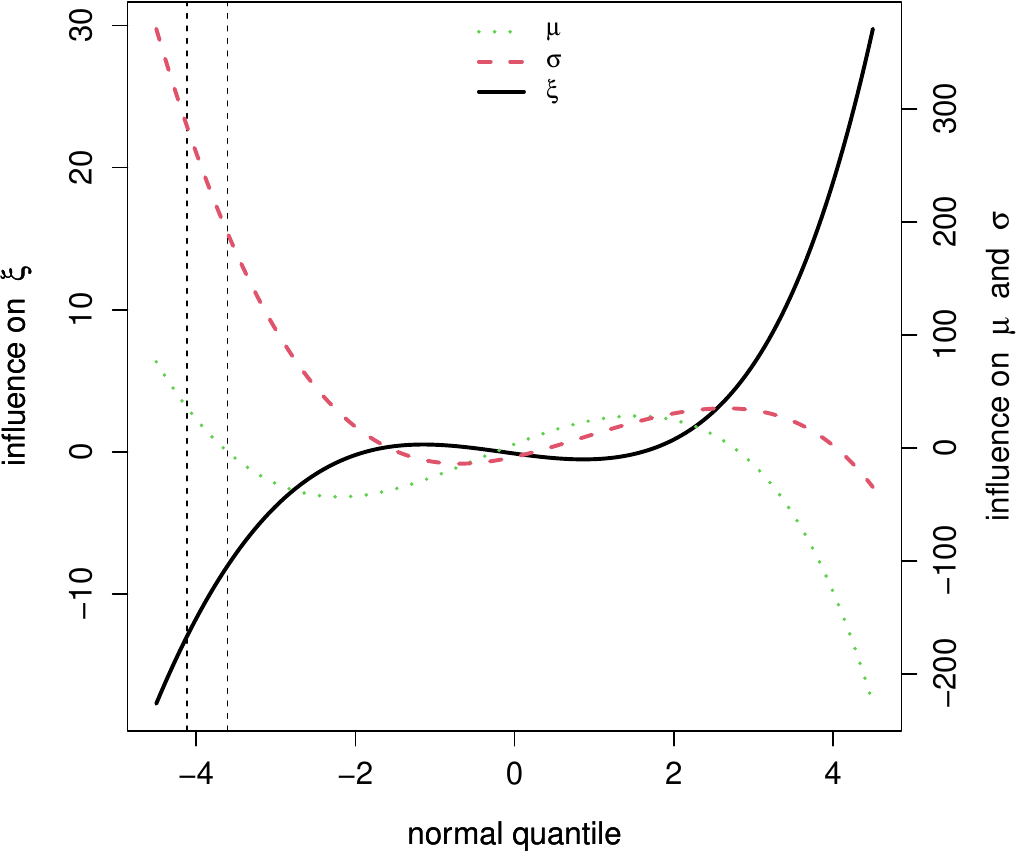}
    \end{subfigure}
    \hfill
    \begin{subfigure}{0.49\textwidth}
    \centering
    \includegraphics[height=7cm]{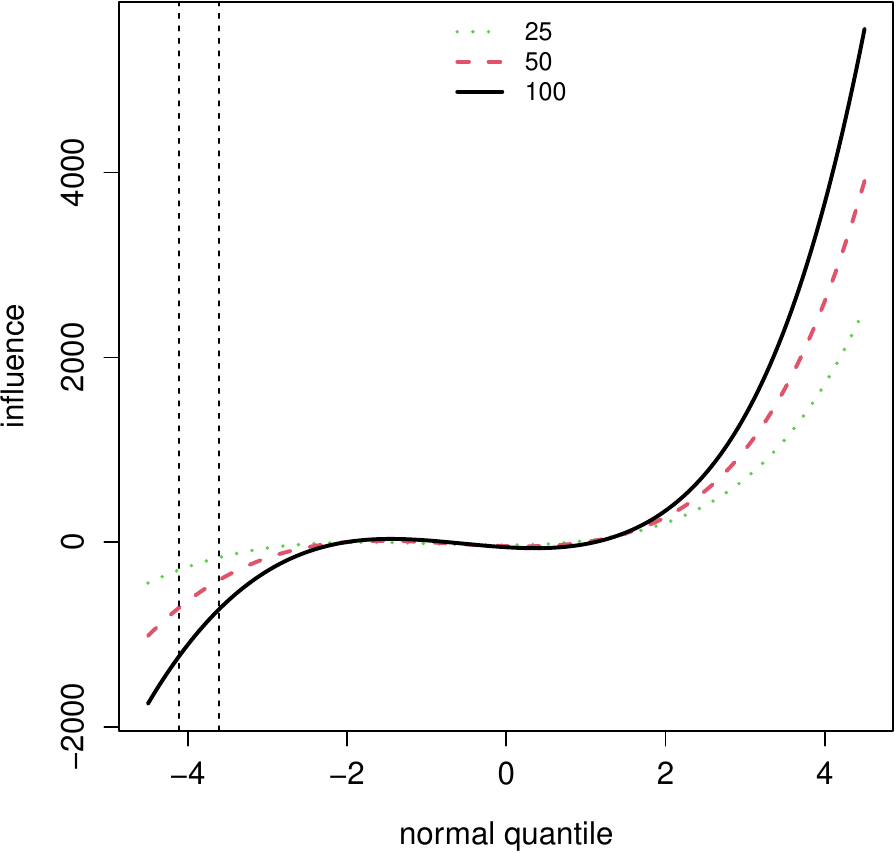}
    \end{subfigure}
    \caption{GEV influence curves for maximum likelihood estimators of $\mu$, $\sigma$ and $\xi$ (left panel) and 25-, 50- and 100-year return levels (right panel) based on the parameter estimates from our approach, plotted against normal quantiles. In each case, the estimated normal quantiles of the two observations highlighted in red in Figure~\ref{fig:ozoneData} of the main paper are indicated by the vertical dashed lines.}
    \label{fig:gev_influence}
\end{figure}

Let $\theta = (\mu, \sigma, \xi)$. The GEV influence function for an observation $y$ is $i_\theta^{-1} {\rm d}\ell(y; \theta)/{\rm d} \theta$, where $\ell(y; \theta)$ is the GEV log-likelihood function and $i_\theta^{-1}$ is the GEV expected information matrix. To aid interpretation, influence functions are expressed on the scale of standard normal quantiles, via $z = \Phi^{-1}\{ G(y; \theta) \}$, where $\Phi$ and $G$ are the distribution functions of a standard normal and GEV($\mu, \sigma, \xi$) distribution, respectively. The left panel of Figure~\ref{fig:gev_influence} shows influence curves for each of the GEV parameters based on the fit to the Plymouth ozone data using our method. Separate vertical scales are used for $\xi$ and $(\mu, \sigma)$ to avoid the curves for $\mu$ and $\sigma$ dominating the plot. The GEV influence function values for $\mu$ and $\sigma$ scale with $\sigma$, and therefore the relatively large influence values in the left panel of Figure~\ref{fig:gev_influence} reflect the size (18.81) of $\hat{\sigma}$. For context, on the scale plotted, the influence function for the sample mean, which is sensitive to a change in the value of an observation, would be the identity function. \cite{DavisonSmith1990} observe that when modelling threshold excesses using a generalised Pareto distribution (GPD), the positive influence of the largest excesses on $\hat{\xi}$ is huge. This is the case in the current context, but here the smallest block maxima also have a strong influence: negative on $\hat{\xi}$ and positive on $\hat{\sigma}$ and, to a lesser extent, on $\hat{\mu}$. Based on the GEV distribution fitted to the Plymouth ozone data using our method, on the standard normal scale the two lowest block maxima have values of approximately $-4$. This explains the differences between the adjusted and unadjusted estimates in Table~\ref{tab:ozoneEstimates}. The right panel of Figure~\ref{fig:gev_influence} shows the corresponding influence curves for the 25-, 50- and 100-year return levels.

\section{Simulation study with correlated data}\label{sm:non-iid}

We include a preliminary investigation of the effects of serial dependence on the estimation of GEV parameters in comparison to the full data case. Raw data are simulated from a max-autoregressive process of order 1, or maxAR(1) process, \citep{DavisResnick1989} where, for $i = 1, 2, \ldots$,  $X_i = \max\{ (1 - \theta) X_{i-1}, \theta Z_i \}$, and $\{Z_i\}$ and $X_0$ have independent unit Fr\'{e}chet distributions. The parameter $\theta \in (0,1]$ is the extremal index, a measure of the degree of local dependence in the extremes of this process. We set $\theta = 0.5$ in our simulation study. The marginal distribution from which the raw data are simulated is unit exponential and the other simulation settings are the same as in the main paper, so the results in Table~\ref{tab:gev_table_maxAR1} may be compared to the exponential case in Table~\ref{tab:gev_table}.

Our adjustment specifies a form of the location and scale parameters of the GEV distribution that accounts for the number $n_i$ of non-missing values in block $i$. In particular, if block $i$ has data missing then $n_i < n$ and $\mu(n_i) < \mu(n) = \mu$ reflects the potential for the observed maximum for block $i$ to be smaller than the maximum that would have been recorded if all data had been observed. A consequence is that our adjustment produces an increase in the estimate, $\hat{\mu}$, of $\mu$ that reflects the potential for the maximum of the unobserved values in a block to be larger than the maximum of the observed values in that block. In the presence of serial dependence, the unobserved values in a block are not independent and their maximum is stochastically smaller than their maximum would be if they were independent. Therefore, we expect that using our adjustment in the presence of non-negligible local dependence in extremes will tend over-compensate, increasing $\hat{\mu}$ relative to the full data case more than is required. 

\newcolumntype{d}[1]{D{.}{.}{#1}}

\begin{table}[!b]
\centering
 \resizebox{0.85\textwidth}{!}{\begin{tabular}{|c|d{2.4}d{2.4}d{2.6}|ccc|ccc|} 
 \hline
& & \multicolumn{1}{c}{bias} & & & sd & & & rmse & \\
approach & \multicolumn{1}{c}{$\mu$} & \multicolumn{1}{c}{$\sigma$} & \multicolumn{1}{c}{$\xi$} \vline & $\mu$ & $\sigma$ & $\xi$ & $\mu$ & $\sigma$ & $\xi$ \\ \hline \hline
adjust & 0.050 & -0.00253 & 0.001206 & 0.036 & 0.037 & 0.037 & 0.062 &
0.037 & 0.037 \\
weight2 & 0.136 & 0.01411 & -0.020468 & 0.062 & 0.052 & 0.049 & 0.149 &
0.054 & 0.053 \\
weight1 & -0.056 & -0.00052 & 0.000081 & 0.034 & 0.037 & 0.038 & 0.065 &
0.037 & 0.038 \\
na\"ive & -0.058 & -0.00056 & 0.000300 & 0.034 & 0.037 & 0.038 & 0.067 &
0.037 & 0.038 \\
discard & -0.016 & -0.02221 & -0.007354 & 0.174 & 0.130 & 0.163 & 0.175
& 0.132 & 0.163 \\ \hline
\end{tabular}}
\caption{Estimation of GEV parameters in comparison to the full data case. The estimated bias, standard deviation (sd) and root mean squared error (rmse) of estimators of $\mu$, $\sigma$ and $\xi$ are given for each approach and for each of the simulation distributions.}
\label{tab:gev_table_maxAR1}
\end{table}

This is indeed what we find here. Whereas our adjusted estimator of $\mu$ was approximately unbiased in Table~\ref{tab:gev_table} of the main paper for the independence case, in Table~\ref{tab:gev_table_maxAR1} its estimated bias is positive. The same phenomenon is observed for all the approaches, that is, the ``weight2'' estimator has much greater positive bias and the other estimators are less negatively biased than in the independence case. Otherwise, the general findings are similar to Table~\ref{tab:gev_table}, suggesting that using our adjustment is still preferable to the other approaches. 

If no raw data are missing, and the data-generating process satisfies a regularity condition, then the extremal index $\theta$ quantifies approximately the effect of this local dependence on the distribution function of block maxima relative to the independence case \citep{Leadbetter1983}. However, if data are missing then this effect depends in a non-trivial way on the locations of the missing values within the block. Even in the unrealistic special case where missing values occur in a regular pattern, such as every second value being missing, even providing bounds for the extremal index of this sub-sampled process is challenging and requires that the underlying process satisfies further conditions \citep{RobinsonTawn2000}.  Therefore, modifying our adjustment to account for local dependence in the extremes of the raw data requires special consideration.

\end{document}